\documentclass[12pt]{article}
\usepackage{amsmath}
\usepackage[colorlinks=true,linkcolor=blue, citecolor=blue]{hyperref}%
\usepackage{amsfonts}
\usepackage{amssymb}
\usepackage{amsthm}
\usepackage{epsfig}
\usepackage{graphicx}
\usepackage[ruled,vlined]{algorithm2e}
\usepackage{hyperref}
\usepackage{subfig}
\usepackage{graphicx}
\usepackage[countmax]{subfloat}
\usepackage{setspace}
\usepackage{subfig}
\usepackage{gensymb}
\usepackage{color}
\usepackage{bm}
\usepackage{authblk}
\usepackage{fullpage}
\usepackage{url}
\usepackage[authoryear]{natbib}
\usepackage{etoolbox}
\usepackage{amsthm}
\usepackage{multirow}
\usepackage[]{algorithm2e}
\usepackage{hyperref}
\usepackage{enumitem}

\newtheorem*{theorem1*}{Proposition 1 (uniform consistency)}
\newtheorem*{theorem2*}{Proposition 2}

\theoremstyle{remark}

\makeatletter

\newif\ifabbreviation
\pretocmd{\thebibliography}{\abbreviationfalse}{}{}
\AtBeginDocument{\abbreviationtrue}

\begin{document}


\title{Analysis of Large Heterogeneous Repairable System Reliability Data with Static System Attributes and Dynamic Sensor Measurement in Big Data Environment}
\author[1]{Xiao Liu} 
\author[2]{Rong Pan} 
\affil[1]{Department of Industrial Engineering\\ University of Arkansas}
\affil[2]{School of Computing, Informatics, Decision Systems Engineering, Arizona State University}

\date{ }

\maketitle

\vspace{0.5cm}
%
%


\singlespacing
\begin{abstract}

In Big Data environment, one pressing challenge facing engineers is to perform reliability analysis for a large fleet of heterogeneous repairable systems with covariates. In addition to static covariates, which include time-invariant system attributes such as nominal operating conditions, geo-locations, etc., the recent advances of sensing technologies have also made it possible to obtain dynamic sensor measurement of system operating and environmental conditions. As a common practice in the Big Data environment, the massive reliability data are typically stored in some distributed storage systems. Leveraging the power of modern statistical learning, this paper investigates a statistical approach which integrates the Random Forests algorithm and the classical data analysis methodologies for repairable system reliability, such as the nonparametric estimator for the Mean Cumulative Function and the parametric models based on the Nonhomogeneous Poisson Process. We show that the proposed approach effectively addresses some common challenges arising from practice, including system heterogeneity, covariate selection, model specification and data locality due to the distributed data storage. The large sample properties as well as the uniform consistency of the proposed estimator is established. Two numerical examples and a case study are presented to illustrate the application of the proposed approach. The strengths of the proposed approach are demonstrated by comparison studies. 

\end{abstract}

\noindent\textbf{Key words:} {\em Repairable Reliability Data Analysis, Recurrence Data, Random Forests, Mean Cumulative Function, Nonhomogeneous Poisson Process.}

\clearpage
\singlespacing
\section{Introduction} \label{sec:one}
Advances in sensing technologies are re-shaping the landscape of statistical analysis for reliability data. For example, a typical repairable system reliability data set contains the failure times from a large fleet of systems as well as a set of covariates associated with each system. Some covariates are static such as 
time-invariant system attributes, while some covariates are dynamic such as sensor measurements of operating and environmental conditions (known as the SOE data). SOE data represent one of the most significant trends in modern reliability analysis in the age of Big Data \citep{Meeker2014, Hong2018}. 
To elaborate the challenges lie ahead, a motivating example is firstly presented.  
\subsection{Motivating Example and Challenges} \label{sec:motivating}
One million oil and natural gas wells are currently operating in the U.S. \citep{EIA2017}. Preventive maintenance of these complex engineering systems is managed by multinational oil and gas service companies which collect and store massive field data sets. Our motivating example is based on a data set which consists of the failure/repair data, system attributes, and sensor measurement of 8232 oil and gas wells installed between 2007 and 2017. 

\begin{figure}%
	\centering
	\subfloat[]{{\includegraphics[width=.45\linewidth]{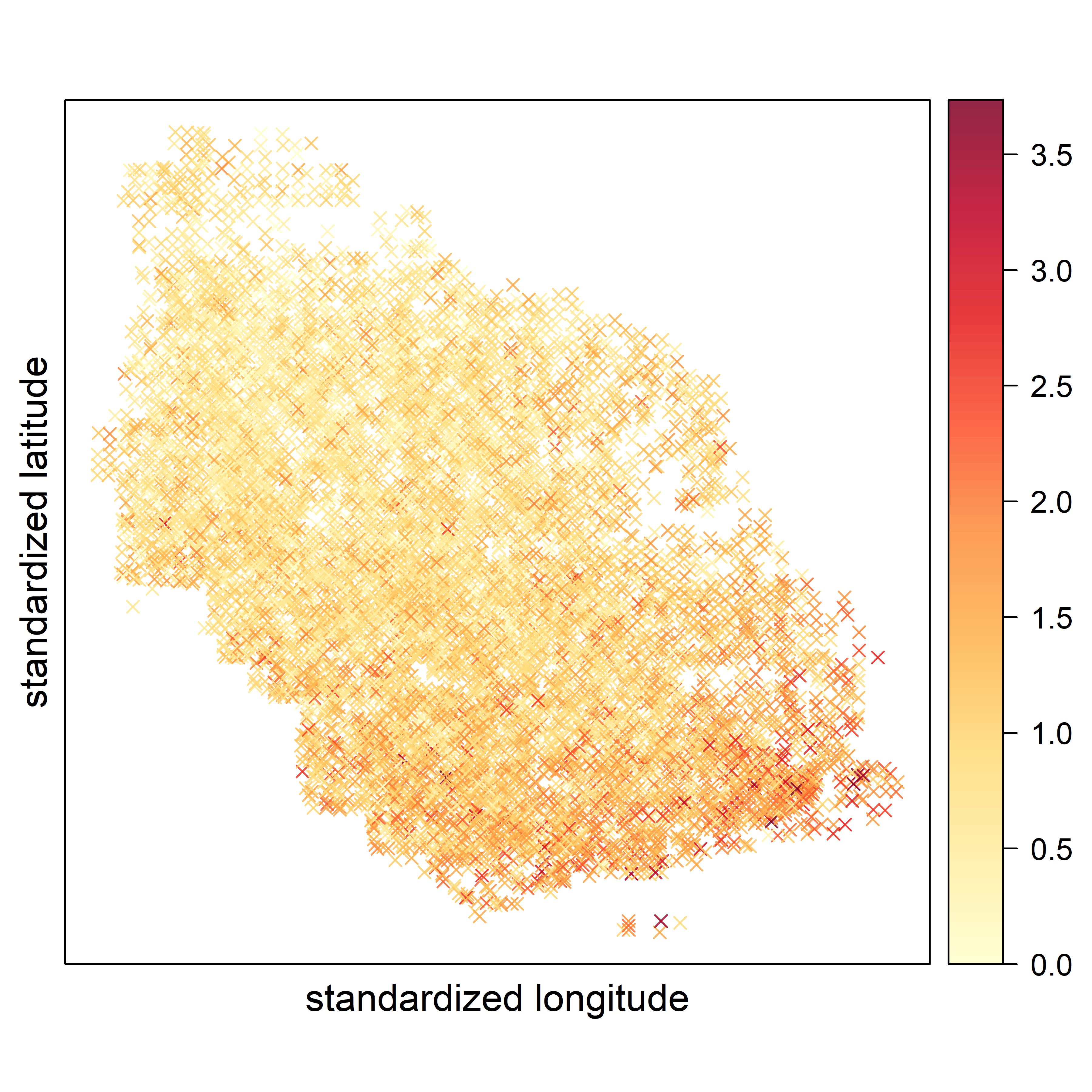} }}%
	\qquad
	\subfloat[]{{\includegraphics[width=.45\linewidth]{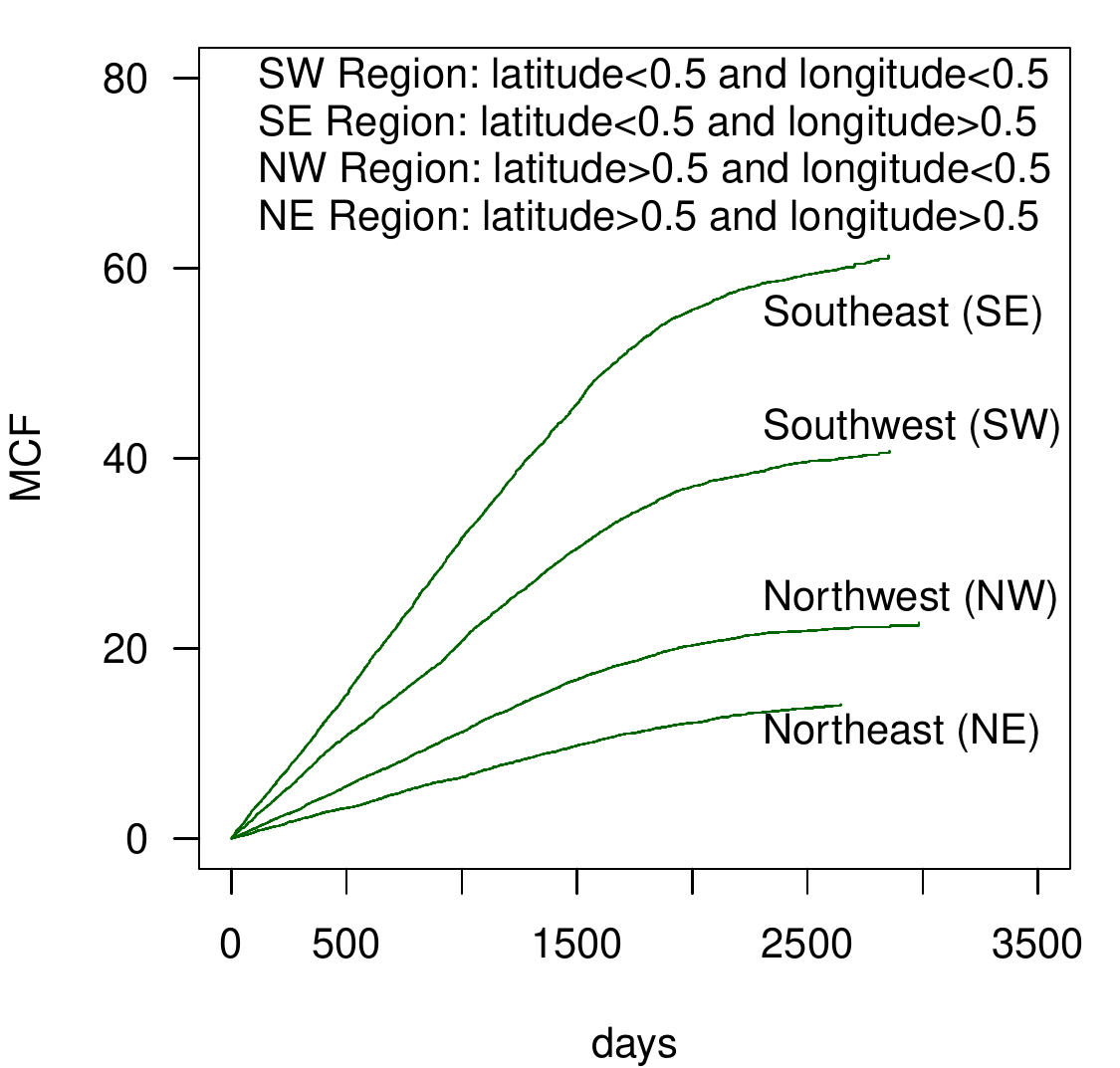} }}%
	\caption{Panel (a) shows the locations of the 8232 wells and the legend indicates the average annual failures; Panel (b) shows the estimated MCF over four geo-regions.}%
	\label{fig:wells}%
\end{figure}

For each well, the ages upon failures are recorded. Figure \ref{fig:wells}(a) shows the geo-locations and the average annual number of failures of these wells. The longitude and latitude are standardized to a unit square $[0,1]^2$ so as to keep the actual well locations confidential. 
Figure \ref{fig:wells}(b) shows the estimated MCF (Mean Cumulative Function) \citep{Nelson1995} for systems over four geo-regions: southwest (SW), southeast (SE), northwest (NW) and northeast (NE). Systems in the southern area appear to have higher overall failure rates than those located in the central and northern regions. Such a spatial trend is often due to system age, soil type, environmental and operating conditions, and others.   

In addition to the failure ages, a number of 15 covariates are available including eight \textit{static} covariates, $x_1 \sim x_8$, and seven \textit{dynamic} covariates, $z_1 \sim z_7$ as follows: 1) (well attributes) covariates $x_1 \sim x_{6}$ are basic well attributes such as well size, nominal working conditions, etc.; 2) (geographical location) covariates $x_7$ and $x_8$ respectively specify the latitude and longitude of each well. Neighboring wells share common environmental conditions such as temperature-humidity variation, soil type, contamination, etc. Although these factors are not directly observed, they may lead to some important spatial patterns of the failure processes; 3) (torque and load) covariates $z_1 \sim z_5$ are related to various dynamic torque and load of each well such as gear torque, load range, etc.; and 4) (stress) covariates $z_{6}$ and $z_{7}$ are respectively the gearbox and structural stress of each well monitored by sensors. 
%
%

\begin{figure}[h!]
	\begin{center}
		\includegraphics[width=1\textwidth]{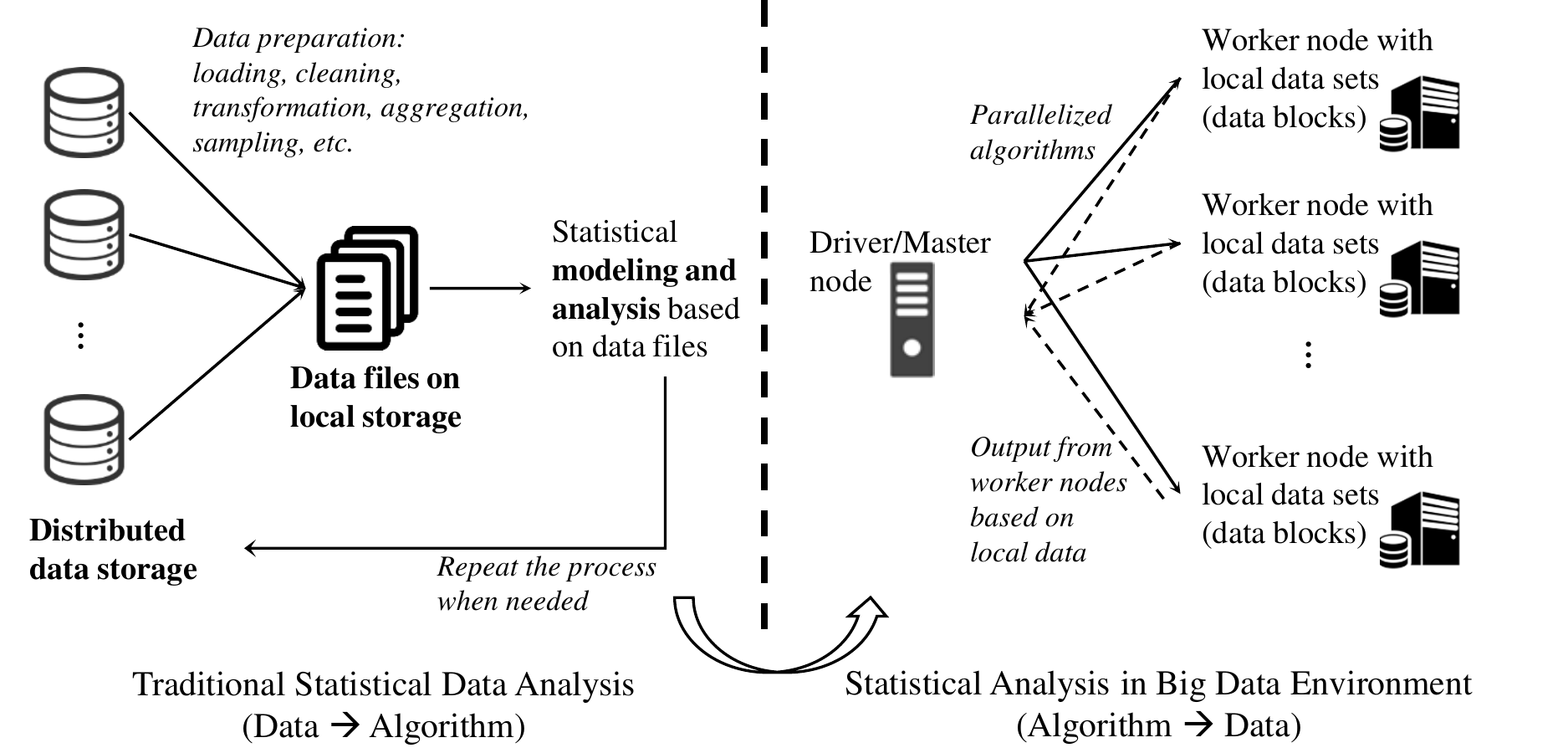}
	\end{center}
	\caption{Traditional data analysis moves data to algorithms, while data analysis in the age of Big Data moves algorithms to data.}
	\label{fig:BigDataAnalytics} 
\end{figure}

To analyze such a repairable system reliability data set, nonparametric graphical methods have been widely used for estimating the MCF which describes the average number of failures for a population of systems up to a certain age \citep{Nelson1995, Meeker1998a}. However, nonparametric graphical methods are less effective in modeling the complex relationship between failure processes and covariates. Parametric point and counting processes have also been widely used to model the recurrence data \citep{Fleming1991, Anderson1993, Meeker1998a, Rigdon2000}. Commonly used models include the Nonhomogeneous Poisson Process (NHPP), renewal process, trend-renewal process, piecewise exponential model, bounded intensity model, etc. \citep{Pulcini2001, Lindqvist2003, Pan2009, Yang2012, Ye2013b, Mittman2018}. 
However, as both the system fleet size and the number of covariates increase, some fundamental challenges arise 
in the modern Big Data analytics environment: 
\begin{itemize}
	[leftmargin=0.1in]
	\setlength\itemsep{0.01em}
	\item  (\textit{Heterogeneity}) It is no longer appropriate to ignore system heterogeneity among a large fleet of field systems \citep{Lindqvist2003, Stocker2007, Ye2013, Xu2017}. In the motivating example, it is impossible to assume that the 8232 wells are from a homogeneous population. Subpopulations always exist due to a number of reasons including the basic system attributes, variation in operating and environmental conditions, maintenance history, etc. Although covariate information may be useful in explaining the system-to-system variation, multiple subpopulations can be governed by fundamentally different failure processes with distinctive dependence structure on covariates. 
	\item (\textit{Model Specification}) It is challenging to force and validate parametric assumptions that adequately describe the link between failure processes and covariates, especially when interactions and nonlinear effects exist among a large number of covariates. This issue is further complicated due to system heterogeneity: both the effects of individual covariates and the relationship between failure processes and covariates can vary across subpopulations. 
	
	\item (\textit{Model Complexity and Interpretability}) Although an increasingly larger number of covariates has been made available due to the advances of sensing technologies, it is evident from our industry practice that many covariates are in fact redundant from either the statistical modeling or domain knowledge perspective. However, as many companies today have invested tremendous resources in collecting and storing data, a common misconception arising from industry is that a statistical model needs to utilize all data provided. 
	Hence, efficient selection of important covariates are particularly important in the Big Data environment, as data-driving models are becoming more complex but seemingly less interpretable. 
	\item (\textit{Data Locality}) In the Big Data environment, the ways data are stored determines how statistical data analysis can be efficiently performed---a critical issue whose impact on reliability analysis has often been overlooked. 
	A common practice in industry today is to store the massive reliability data on distributed storage such as the Hadoop distributed file system. This leads to a critical difference between traditional statistical analysis and statistical analysis in the age of Big Data: traditional data analysis moves \textit{data to algorithms}, while data analysis in the age of Big Data moves \textit{algorithms to data}; as shown in Figure \ref{fig:BigDataAnalytics}.
	
\end{itemize}

To elaborate, traditional data analysis often starts with data preparation which pulls data from a distributed storage with multiple servers. Then, through a series of data pre-processing,
data files (such as text files and Excel spreadsheet) are generated and saved to a local storage. Only after such a time-consuming process can the modeling and statistical data analysis be performed based on the data files generated. Whenever there is an update on raw data sets in the distributed storage, or, there are new requirements arising from the modeling point of view, the above lengthy data preparation process is repeated which typically involves the transfer of a huge amount of data. 
Data analysis in the age of Big Data works in exactly the opposite direction. Instead of pulling data from different servers, the statistical analysis procedures are assigned to different servers where data are stored in a distributed manner. This is known as ``\textit{Data Locality}'' which has tremendous advantages in terms of reliability, scalability, efficiency and security \citep{Spark2018}. The well-known MapReduce programming model---originally referred to the proprietary Google technology---is based on such an idea that any algorithms can be executed on smaller subsets of a larger data set, and the outputs generated from smaller subsets are then merged to form the final results. Through the idea of MapReduce, computations are parallelized on local nodes which is critical when working with large data. 



\subsection{Overview and Literature Review}
To tackle the challenges discussed above, this paper investigates an approach which integrates two powerful methods respectively from the modern statistical learning and classical repairable system reliability analysis, i.e., the Random Forests (RF) algorithm \citep{Breiman2001} and the nonparametric MCF estimator \citep{Nelson1995}. The RF method, which constructs ensembles (forests) from de-correlated base learners (trees), is one of the most popular ensemble methods which has gained tremendous success in practice. The randomness is introduced into the RF through two mechanisms: 1) each tree is grown based on a bootstrap sample (bagging), and 2) at each node of a tree, the optimum splitting variable and split point are chosen from a randomly sampled candidate variable set. The nonparametric MCF estimator, on the other hand, is one of the most successful methods for analyzing repairable system reliability data (i.e., recurrence data) in industry \citep{Nelson1995, Meeker1998a, Doganaksoy2006, Zuo2012}. 
The theory behind the nonparametric MCF estimator is deeply rooted in point processes \citep{Fleming1991}. 

In the literature, tree-based methods for analyzing survival data (i.e., time-to-event data) with censoring have attracted much attention. 
A sum-of-trees (ensemble) model is essentially an additive model with multivariate components that effectively handle the complex interaction effects among covariates \citep{Chipman2010, Pratola2016}.
For right-censored data, \cite{Hothorn2006} introduced an RF algorithm and a generic gradient boosting algorithm for predicting the survival times of patients suffering from acute myeloid leukemia based on clinical and genetic covariates; also see \cite{Hothorn2004} for bagging survival trees. \cite{Ishwaran2008} proposed the Random Survival Forests (RSF) algorithm that integrates the classical Kaplan-Meier estimator into the framework of RF. \cite{Fan2006, Fan2009} investigated trees for correlated and multivariate survival data. \cite{Bacchetti1995} proposed an approach which decomposes a subject into multiple pseudo-subjects, and the pseudo-subjects can be split across many nodes as a function of time. \cite{BouHamad2009} incorporated dynamic covariates in the tree-based survival analysis by allowing subjects to be split across different nodes depending on the time windows. A review of tree-based methods for analyzing time-to-failure data can be found in \cite{BouHamad2011}.

This work focuses on the analysis of large recurrence data with both static and dynamic covariates arising from repairable systems. We investigate the statistical properties of the estimator and show how the proposed approach naturally fits into the modern Big Data environment. To our best knowledge, such work has not yet been done in the literature. 
We organize the paper as follows: Section \ref{sec:overview} presents the proposed algorithm by only considering static covariates. Technical details and some useful theoretical results (i.e., consistency and large-sample variance) are presented. In Section \ref{sec:extension}, we extend the framework to handle both static and dynamic covariates. Section \ref{sec:numerical} presents two numerical examples to demonstrate the advantages of the proposed method over some conventional methods. Section \ref{sec:case} revisits the motivating example and illustrates the application of the proposed approach.

\section{The RF-R Algorithm} \label{sec:overview}
\subsection{Overview} \label{sec:algorithm}
Consider a typical repairable system reliability data set which consists of $n$ systems. For any system $i$ $(i=1,2,...,n)$, we observe a vector $\bm{y}_i=(y_{i,1},...,y_{i,r_i},c_{i})$ which contains a sequence of $r_i$ failure ages, $y_{i,1},...,y_{i,r_i}$, and the right censoring time $c_{i}$. Associated with system $i$ there exists a number of $p$ attributes, $\bm{x}_i=(x_{i,1},x_{i,2},...,x_{i,p})$. Without loss of generality, it is assumed that all covariates are standardized on the unit interval $[0,1]$, and the covariate space is hence a multi-dimensional cube $[0,1]^p$. Let $\Lambda(t)$ denote the number of failures in the time interval $(0,t]$, then, the expectation of $\Lambda(t)$ is the MCF which characterizes the reliability of a population of repairable systems \citep{Nelson1995, Meeker1998a}.

\singlespacing
\begin{algorithm}[H]
	
	\KwData{ ($\bm{x}_i,\bm{y}_i$) for $i=1,2,...,n$}
	
	\textbf{Step 1 }Draw $B$ bootstrap samples from the original data.
	
	\textbf{Step 2 }
	\For{b=1,...,B}{
		Grow a random MCF tree by recursively repeating the following steps at each node of the tree:
		
		\textbf{2.1)} Select $m$ covariates at random from the $p$ covariates.
		
		\textbf{2.2)} Pick the best covariate and its split point among the $m$ selected covariates by maximizing the difference of the MCF between daughter nodes.
		
		\textbf{2.3)} Grow the tree to full size under the condition that each terminal node contains at least $d_0$ systems with at least one failure.
	}
	
	\textbf{Step 3 }Obtain the ensemble MCF by averaging the MCF obtained from all trees.
	
	\textbf{Step 4 }Compute the Out-of-Bag (OOB) prediction error for the ensemble MCF.
	\caption{The RF-R algorithm}
\end{algorithm}
\singlespacing
To tackle the challenges described in Section \ref{sec:motivating}, an algorithm, called RF-R (\textbf{R}andom \textbf{F}orests for \textbf{R}epairable System Reliability Analysis), is proposed to integrate the nonparametric MCF estimator into the framework of RF. The RF-R algorithm is described in Algorithm 1, while the technical and theoretical details are presented in Sections \ref{sec:details} and \ref{sec:theory}.

The RF-R integrates the nonparametric MCF estimator into the framework of RF, and hence has the same computational complexity as RF, i.e., $\mathcal{O}(m n \log n)$. The choice of $m$ is case-dependent; for example, $m=\left \lfloor p/3 \right \rfloor$ as recommended in \cite{Hastie2009}. By integrating the nonparametric MCF estimator with RF, the RF-R algorithm achieves the following advantages: 1) (heterogeneity) for each bootstrap sample, the algorithm uses a binary tree to divide the covariate space into a set of rectangular partitions represented by the terminal nodes of a tree. In other words, heterogeneous systems are divided into sub-groups based on their attributes. Then, separate reliability models, such as the MCF in this case, are constructed to characterize the system reliability on different terminal nodes; 2) (model specification) the nonparametric MCF estimator enables us to avoid the difficult specification of the complicated relationship between failure processes and covariates. Complex interaction structures in data can also be automatically captured by the tree-based methods \citep{Chipman2010}. The sample size in the Big Data environment is large enough to overcome the traditional limitations of nonparametric methods; and 3) (data locality) the RF-R algorithm, which can be implemented in a modern distributed computing environment, does not require any transfer of the original data sets across the worker nodes and satisfies the basic requirements for Big Data analytics as discussed in Section \ref{sec:motivating}. 

To elaborate, Step 1 involves generating bootstrap samples on local \textit{worker} nodes. This process can be done in parallel across worker nodes. In Step 2, a MCF tree is grown which involves recursively splitting the tree nodes until the termination condition is met. For each splitting, the \textit{driver} node selects $m$ covariates at random from the $p$ covariates, and creates the candidate split point for each of the selected $m$ covariates. 
After receiving the selected covariates and the candidate split points, the \textit{worker} node, in parallel, performs the splitting of the local bootstrap data into two sub-populations, and constructs the (local) MCF respectively for both sub-populations. The constructed MCF, using the data stored on the worker node, are returned to the \textit{driver} node. After that, the \textit{driver} node merges the local MCF constructed from all worker nodes using the method to be discussed in \ref{sec:mcf}. Finally, the optimum splitting covariate and split-point are found on the \textit{driver} node. The tree structure is updated on the \textit{driver} node until the algorithm terminates. 
\subsection{Technical Details}\label{sec:details}

Let $T(\bm{x};\Theta^{(b)})$ be a binary MCF tree based on the bootstrap sample $b$, $b=1,2,...,B$. Here, $\bm{x}$ is a vector of covariates and the parameter, $\Theta^{(b)}$, fully specifies the tree in terms of splitting covariates, split points, and the constructed MCF on terminal nodes. 
Let $(h^{(b)}_1, h^{(b)}_2, ..., h^{(b)}_M)$ be a set of $M$ terminal nodes of the $b$th tree and each terminal node defines a binary partition on the covariate space. Then, for a vector of covariates $\bm{x} \in h^{(b)}_j$, the MCF is given by:
\begin{equation} \label{eq:mcf-tree}
   \widehat{\mathrm{MCF}}^{(b)}(t;\bm{x}) = \sum_{j=1}^{M} \widehat{\mathrm{MCF}}_{h^{(b)}_j}(t)I\{\bm{x} \in h^{(b)}_j\}
\end{equation}
where $\widehat{\mathrm{MCF}}_{h^{(b)}_j}(t)$ is the MCF constructed on terminal node $h^{(b)}_j$.
The bootstrap ensemble MCF is computed by averaging over the $B$ trees:
\begin{equation} \label{eq:mcf-ensemble}
\widehat{\mathrm{MCF}}^{(*)}(t;\bm{x}) = \frac{1}{B}\sum_{b=1}^{B}\widehat{\mathrm{MCF}}^{(b)}(t;\bm{x})=   \frac{1}{B}\sum_{b=1}^{B}\sum_{j=1}^{M} \widehat{\mathrm{MCF}}_{h^{(b)}_j}(t)I\{\bm{x} \in h^{(b)}_j\}.
\end{equation}

The greedy algorithm can be used in step 2 of Algorithm 1 which recursively decides on the optimum splitting covariates, split points, and the tree topology \citep{Hastie2009}. Given the candidate splitting covariate $j$ and the split point $\tilde{x}_j$, an intermediate node $\tilde{h}$ can be split into two daughter nodes. The greedy algorithm seeks the optimum splitting covariate and the split point that maximize the difference between the two MCF, $\widehat{\mathrm{MCF}}_{\tilde{h}}^{(-)}(t)$ and $\widehat{\mathrm{MCF}}_{\tilde{h}}^{(+)}(t)$, respectively constructed on the left and right daughter nodes:
\begin{equation} \label{eq:split}
\max_{j,\tilde{x}_j} g(\widehat{\mathrm{MCF}}_{\tilde{h}}^{(-)}(t), \widehat{\mathrm{MCF}}_{\tilde{h}}^{(+)}(t) )
\end{equation}
where $g$ is some distance measure between two MCF. Hence, two fundamental questions need to be addressed: the construction of MCF on a given tree node for distributed data (Section \ref{sec:mcf}), and the choice of the splitting rule in (Section \ref{sec:split}). 

\subsubsection{MCF for Large Distributed Data Sets} \label{sec:mcf}
For distributed data sets, the MCF needs to be constructed in a distributed manner such that each worker node constructs the MCF using the local data on that worker node and the driver node merges the MCF constructed from all worker nodes. By modifying algorithm 16.1 in \cite{Meeker1998a}, we obtain the following algorithm which constructs MCF on a tree node using distributed data on worker nodes. This approach can be used to obtain the MCF on both intermediate and terminal nodes, and the subscript $h$ in (\ref{eq:split}) is suppressed. 

\begin{algorithm}
	\KwData{Prepare the data set $\mathfrak{D}_w$ on each local worker node $w$ ($w=1,2,...,W$) from which a MCF is to be constructed.} 
	\textbf{Step 1 (Map phase)} For each worker node $w$, the following steps are executed in parallel, i.e.,
	\For{w=1,...,W}{
		
		\textbf{1.1.} Find the ordered unique failure times, $t_{1}<t_{2}<...<t_{n^{(w)}}$, where $n^{(w)}$ is the number of unique failure times in $\mathfrak{D}_w$.
		
		\textbf{1.2.} Compute the number of failure $d_i(t_{k})$ for system $i$ at time $t_{k}$ ($k=1,2,...,n^{(w)}$) in $\mathfrak{D}_w$. 
		
		\textbf{1.3.} Let $\delta_i(t_{k})=1$ if system $i$ is still being observed at $t_{k}$; $\delta_i(t_{k})=0$ otherwise.
		
		\textbf{1.4.} Compute the local MCF from the data set $\mathfrak{D}_w$ for $j=1,2,...,n^{(w)}$ 
		\begin{equation} \label{eq:mcf1}
		\widehat{\mathrm{mcf}}^{(w)}(t_{j}) = \sum_{k=1}^{j} \left\{  \frac{ \sum_{i=1}^{n^{(w)}} \delta_i(t) d_i(t_k) }{\sum_{i=1}^{n^{(w)}} \delta_i(t_k)}   \right\} = \sum_{k=1}^{j} \frac{d.^{(w)}(t_k)}{\delta.^{(w)}(t_k)} = \sum_{k=1}^{j} \bar{d}^{(w)}(t_k)
		\end{equation}
		where the estimate, $\widehat{\mathrm{mcf}}^{(w)}(t)$, is a step function, with jumps at failure times, but constant between the failure times.
		
	}
	
	\textbf{Step 2 (Reduce phase)}
	Merge the local MCF on the driver node:
	\begin{equation} \label{eq:mcf2}
	\widehat{\mathrm{MCF}}(t) = \frac{\sum_{m=1}^{W} \widehat{\mathrm{mcf}}^{(w)}(t)}{W}
	\end{equation}
	
	\caption{Nonparametric MCF for Distributed Data Sets}
\end{algorithm}

From \cite{Nelson1995} and \cite{Meeker1998a}, $\widehat{\mathrm{mcf}}^{(w)}(t_{j})$ in (\ref{eq:mcf1}) is unbiased for the data on the worker node $w$ and has the large-sample approximate variance as follows:
\begin{equation} \label{eq:var1}
\widehat{\mathrm{Var}}(\widehat{\mathrm{mcf}}^{(w)}(t)) = \sum_{i=1}^{r^{(w)}} \left\{ \sum_{k=1}^{j} \frac{\delta_i(t_k)}{\delta.^{(b)}(t_k)}\left( d_i(t_k) - \bar{d}^{(b)}(t_k) \right ) \right\}^2.
\end{equation}

Hence, assuming that individual systems are independent conditioning on the covariates, $\widehat{\mathrm{MCF}}(t)$ in (\ref{eq:mcf2}) is also unbiased with its variance estimated by:
\begin{equation} \label{eq:var2}
\widehat{\mathrm{Var}}( \widehat{\mathrm{MCF}}(t) )= \frac{\sum_{m=1}^{W} \widehat{\mathrm{Var}}( \widehat{\mathrm{mcf}}^{(w)}(t))}{W^2}.
\end{equation}

It is also possible to obtain the covariance of $\widehat{\mathrm{MCF}}(t_j)$ and $\widehat{\mathrm{MCF}}(t_p)$:
\begin{equation} \label{eq:cov1}
\begin{split}
\mathrm{Cov}( \widehat{\mathrm{MCF}}(t_j), \widehat{\mathrm{MCF}}(t_p)) & = \frac{1}{W^2} \mathrm{Cov}\left( \sum_{w=1}^{W}\sum_{k=1}^{j} \bar{d}^{(w)}(t_k), \sum_{w=1}^{W}\sum_{k=1}^{p} \bar{d}^{(w)}(t_k)\right) \\ & = \mathrm{Cov}( \widehat{\mathrm{mcf}}^{(w)}(t_j), \widehat{\mathrm{mcf}}^{(w)}(t_p)).
\end{split}
\end{equation}

Here, $\mathrm{Cov}( \widehat{\mathrm{mcf}}^{(w)}(t_j), \widehat{\mathrm{mcf}}^{(w)}(t_p)) = \sum_{k=1}^{j}\sum_{k'=1}^{p} \mathrm{Cov} (\bar{d}^{(w)}(t_k),\bar{d}^{(w)}(t_{k'}))$
and
\begin{equation} \label{eq:cov3}
\mathrm{Cov} (\bar{d}^{(w)}(t_k),\bar{d}^{(w)}(t_{k'})) =
\begin{cases}
\frac{\mathrm{Cov}(d^{(w)}(t_k),d^{(w)}(t_{k'}))}{\delta.^{(w)}(t_k)} & k<k' \\ \frac{\mathrm{Cov}(d^{(w)}(t_k),d^{(w)}(t_{k'}))}{\delta.^{(w)}(t_{k'})} & k>k' \\
\frac{\mathrm{Var}(d^{(w)}(t_k))}{\delta.^{(w)}(t_{k})} & k=k'
\end{cases}
\end{equation}
where $\mathrm{Var}(d^{(w)}(t_k))$ and $\mathrm{Cov}(d^{(w)}(t_k),d^{(w)}(t_{k'}))$ are estimated by \citep{Meeker1998a}:
\begin{equation} \label{eq:cov4}
 \widehat{\mathrm{Var}}(d^{(w)}(t_k))= \sum_{i=1}^{n^{(w)}}\frac{\delta_i(t_k)}{\delta.^{(w)}(t_k)}(d_i(t_k)-\bar{d}^{(w)}(t_k))^2, \quad t_k<t_{k'},
\end{equation}
\begin{equation} \label{eq:cov5}
\widehat{\mathrm{Cov}}(d^{(w)}(t_k),d^{(w)}(t_{k'})) = \sum_{i=1}^{n^{(w)}}\frac{\delta_i(t_{k'})}{\delta.^{(w)}(t_{k'})}(d_i(t_{k'})-\bar{d}^{(w)}(t_{k'}))d_i(t_{k}), \quad t_k<t_{k'}.
\end{equation}


\subsubsection{The Splitting Rule} \label{sec:split}
The splitting rule for a node maximizes the difference between the MCF constructed on the two daughter nodes; see (\ref{eq:split}). 
The RF-R algorithm adopts the following splitting approaches.
\begin{itemize}
   [leftmargin=0.1in]
	\setlength\itemsep{0.05em}
   \item (Splitting based on the log-rank test). For any parent node which contains a number of $k$ pooled failure times, $t_1<t_2<...<t_k$, let
   \begin{equation} \label{eq:Z}
   Z(t_i) = \widehat{\mathrm{MCF}}^{(-)}(t_i) - \widehat{\mathrm{MCF}}^{(+)}(t_i), \quad i=1,2,...,k
   \end{equation}
   be the difference of the MCF estimates at time $t_i$ between the left and right daughter nodes.
   It follows from (\ref{eq:var2}) that $\mathrm{Var}(Z(t_i)) = \mathrm{Var}(\widehat{\mathrm{MCF}}^{(-)}(t_i))+\mathrm{Var}(\widehat{\mathrm{MCF}}^{(+)}(t_i))$ and $\mathrm{Cov} (Z(t_i),Z(t_j)) =  \mathrm{Cov} ( \widehat{\mathrm{MCF}}^{(-)}(t_i), \widehat{\mathrm{MCF}}^{(-)}(t_j)  )  + \mathrm{Cov}(  \widehat{\mathrm{MCF}}^{(+)}(t_i), \widehat{\mathrm{MCF}}^{(+)}(t_j)   )$ for $i,j=1,2,...,k$ and $i \neq j$.
   Let $\bm{\Sigma}$ be the covariance matrix of $\bm{Z}=(Z(t_1),Z(t_2),...,Z(t_k))$, the test statistic is given by the quadratic form $\bm{Z} \bm{\Sigma} \bm{Z}^T$ \citep{Klein2005}.
   Since the MCF estimate is asymptotically normal \citep{Meeker1998a}, the test statistic has a chi-squared distribution for large samples with $k-1$ degree of freedom, under the null hypothesis that $\widehat{\mathrm{MCF}}^{(-)}(t)=\widehat{\mathrm{MCF}}^{(+)}(t)$.
   
   \item (Splitting based on the $L^2$ distance). The optimum splitting essentially maximizes some distance between $\widehat{\mathrm{MCF}}^{(-)}(t)$ and $\widehat{\mathrm{MCF}}^{(+)}(t)$ on the two daughter nodes. Hence, a natural distance measure between two real-valued functions in the $L^2$ space is given by:
   \begin{equation} \label{eq:inner}
   ||  \widehat{\mathrm{MCF}}^{(-)}(t)-\widehat{\mathrm{MCF}}^{(+)}(t) ||_2  = \left( \int_{0}^{t_k}\left(\widehat{\mathrm{MCF}}^{(-)}(t)-\widehat{\mathrm{MCF}}^{(+)}(t)\right)^2dt \right)^{\frac{1}{2}} 
    \approx \left( \sum_{i=1}^{k}Z^2(t_i)\right)^{\frac{1}{2}}.
   \end{equation}
   The square root of the log-rank test statistic reduces to (\ref{eq:inner}) when $\bm{\Sigma}$ becomes an identify matrix. Our numerical experiment shows that the two rules generate comparable performance in terms of accuracy. However, the splitting based on the $L^2$ distance is much faster as it does not require the evaluation of the covariance matrix in the log-rank test statistic. 
\end{itemize}
\subsubsection{Out-of-Bag (OOB) Error} \label{sec:oob}
The use of OOB samples is an important feature under the framework of RF. The MCF of an OOB system is constructed by only averaging those trees in which this particular system does not appear, thus the OOB prediction error is close to that of a cross-validation. The stabilization of the OOB error also helps to determine the number of trees in a forest. 

Let $\gamma_i(b)=1$ if system $i$ is not contained in the bootstrap sample $b$, otherwise, $\gamma_i(b)=0$. Then, the OOB ensemble MCF for system $i$ can be calculated as:
\begin{equation} \label{eq:mcf-oob}
\widehat{\mathrm{MCF}}^{(\mathrm{OOB})}(t;\bm{x}) = \frac{\sum_{b=1}^{B}\gamma_i(b)\sum_{j=1}^{M} \widehat{\mathrm{MCF}}_{h^{(b)}_j}(t)I\{\bm{x} \in h^{(b)}_j\}}{\sum_{b=1}^{B}\gamma_i(b)}.
\end{equation}

The quantification of prediction error depends on specific applications. For example, if the MCF at a particular time is of interest, the prediction error can be measured by the Mean Squared Error of the predicted MCF at that time. In this paper, we follow the idea of \cite{Ishwaran2008} and choose a more general error measure, known as the Harrell's concordance index (or, C-index). The C-index was firstly proposed in \cite{Harrell1982} for evaluating the amount of information a medical test provides about individual patients; also see \cite{Brentnall2018}. For the problem considered in this paper, the C-index can be interpreted as the empirical probability of correctly ranking any two systems in terms of their reliability, and can be calculated in the following way: 1) form all pairs of systems from the OOB sample, and the total number of system pairs is ${n^\text{OOB}}\choose{2}$ with $n^{\text{OOB}}=\sum_{b=1}^{B}\gamma_i(b)$ being the total number of OOB samples; 2) for each pair $i$, rank the two systems based on some reliability measures, such as the mean intensity, the cumulative number of failures up to a given time, etc. These reliability measures can be calculated from the observed failure times; 3) for each pair $i$, rank the two systems based on the same reliability measure, which is calculated from the predicted MCF. If the predicted rankings are consistent with the observed rankings, let $C_i=1$ for pair $i$, otherwise  $C_i=0$; and 4) the C-index for an OOB sample is given by ${{\text{OOB}}\choose{2}}^{-1}\sum C_i$.
%
\subsection{Theoretical Results} \label{sec:theory}
Some useful properties of the RF-R estimator can be obtained by extending the existing work \citep{Fleming1991, Anderson1993, Ishwaran2010}. As discussed in Section \ref{sec:overview}, this paper considers the standardized covariate space $\mathbb{X}$ which is a multi-dimensional cube $[0,1]^p$. 
\cite{Ishwaran2010} showed the uniform consistency of RSF for right-censored time-to-failure data under the assumption that the covariate space $\mathbb{X}$ has finite cardinality and the covariate $\bm{X}$ for any system is randomly sampled from a discrete covariate space $\mathbb{X}$ according to the marginal distribution $\mu$, i.e., $\mu(A)=\mathbb{P}(\bm{X} \in A)$ for some subset $A$ of $\mathbb{X}$. Adopting the same assumption and for any covariate $j$, $j=1,2,...,p$, the continuous covariate space $[0,1]$ is discretized to a large number of $L_j$ equal-width bins in order to ensure high granularity. In numerical computation, even if the covariate is continuous, discretizing covariates is always needed when splitting a tree node.
The proposition below shows the uniform consistency of the ensemble RF-R estimator.
\begin{theorem1*}
Let $v(\cdot;\bm{x})$ be the strictly positive recurrence rate given the covariate $\bm{x}$, and let $\tau=\min(\tau(\bm{x}):\bm{x} \in \mathbb{X} )$ with $\tau(\bm{x}) = \mathrm{sup} \{t: \int_{0}^{t}v(s;\bm{x})ds < \infty \}$. If $\mathbb{P}\{C>c\}>0$ for $c\in [0,\tau)$, then, for $t\in (0,\tau)$ and all $\epsilon>0$ we have
\begin{equation} \label{eq:consistency}
\lim_{N\rightarrow \infty} \mathbb{P}\left\{  \sup_{s\in [0,t]} \left | \mathbb{E}_{\bm{X}}(\widehat{\mathrm{MCF}}^{(*)}(s;\bm{X})) - \mathbb{E}_{\bm{X}}(\mathrm{MCF}(s;\bm{X})) \right | > \epsilon \right\} = 0.
\end{equation}
\end{theorem1*}

The proof is provided in the Appendix. 
Recall that, the fundamental idea behind RF is to reduce the variance through bagging and de-correlating trees. For a given covariate $\bm{x}$, it is well-known that the variance of the ensemble trees is given by \citep{Hastie2009}
\begin{equation} 
\rho(\bm{x})\mathrm{Var}(\widehat{\mathrm{MCF}}(t;\bm{x})) + B^{-1}(1-\rho(\bm{x}))\mathrm{Var}(\widehat{\mathrm{MCF}}(t;\bm{x}))
\nonumber
\end{equation}
where $\mathrm{Var}(\widehat{\mathrm{MCF}}(t;\bm{x}))$ is the sampling variance of any single randomly drawn tree, and $\rho(\bm{x})$ is the correlation between a pair of randomly chosen trees for a given covariate $\bm{x}$. Hence, when $B\rightarrow \infty$, the variance of the ensemble trees becomes $\rho(\bm{x})\mathrm{Var}(\widehat{\mathrm{MCF}}(t;\bm{x}))$.

In \cite{Hastie2009} (page 599), the authors provided a simple approach which allows us to numerically investigate the correlation $\rho(\bm{x})$ between pairs of trees drawn from a forest, while Proposition 2 below gives the asymptotic variance of a single randomly drawn tree.
\begin{theorem2*}
	Let $\delta_{\bm{x}}(t)=1$ if there is at least one system which is still under observation at time $t$ with covariate $\bm{x}$, and $\delta_{\bm{x}}(t)=0$ otherwise, and let $\widetilde{\mathrm{MCF}}(t;\bm{x}) = \int_{0}^{t}I\{\delta_{\bm{x}}(t)>0\}d\mathrm{MCF}(t;\bm{x})$ and $\widetilde{\mathrm{MCF}}(t;\bm{X})=\sum_{x\in\mathbb{X}}\widetilde{\mathrm{MCF}}(t;\bm{x})$.
	Then, for $t\in (0,\tau)$ with $\tau$ being defined in Proposition 1, the asymptotic variance of a randomly drawn tree $\widehat{\mathrm{MCF}}(t;\bm{X})$ is given by:
	\begin{equation}
	\label{eq:asym_var}
	\mathrm{Var}\left( \sqrt{n} (\widehat{\mathrm{MCF}}(t;\bm{X}) - \widetilde{\mathrm{MCF}}(t;\bm{X})) \right) = \sum_{x} \phi(\bm{x},t)
	\end{equation}  
	where $\phi(\bm{x},t) = \int_{0}^{t}\pi^{-1}(s)(1-\Delta \mathrm{MCF}(s;\bm{x}))d\mathrm{MCF}(s;\bm{x}))$,
	and $\pi(\cdot)$ is the density function of the random censoring time, and $\Delta \mathrm{MCF}(t;\bm{x}) \equiv \mathrm{MCF}(s;\bm{x}) - \lim_{s\uparrow t}\mathrm{MCF}(s;\bm{x})$.
\end{theorem2*}
The proof is available in the Appendix. 

\section{Extended RF-R to Incorporate Dynamic Covariates} \label{sec:extension}
In this section, we show that the RF-R algorithm can be extended to handle both static system attributes and dynamic sensor measurements. For any system $i$, in addition to the event times $\bm{y}_i$ and a vector of static covariates $\bm{x}_i$, we also observe a $q$-dimensional time series, $\bm{z}_i(t)=(z_{i,1}(t), z_{i,2}(t), ...,z_{i,q}(t))$, where $z_{i,j}(t)$ can be the $j$th sensor measurement of some operating and environmental condition at time $t$. 

Incorporating dynamic covariates into a tree-based method is challenging. 
As shown in Section \ref{sec:case}, systems in the field can experience distinctive operating and environmental conditions, which typically have cumulative effects on the failure process. 
Hence, we extend the RF-R algorithm to include dynamic covariates based on the following strategy: the split of a tree node is based on the static system attributes, while the data on each node are modeled by a parametric model incorporating the dynamic covariates. 
In other words, we consider the scenario where the subpopulations among a large fleet of systems are mainly characterized by the static system attributes, while the dynamic sensor measurement is used for explaining the system-to-system variation of the failure processes among systems sharing similar attributes. 
In particular, for a vector of covariates $\bm{x} \in h^{(b)}_j$, the corresponding intensity function of the failure process is given by:
\begin{equation} \label{eq:lambda-tree}
\widehat{\lambda}^{(b)}(t;\bm{x}) = \sum_{j=1}^{M} \widehat{\lambda}_{h^{(b)}_j}(t)I\{\bm{x} \in h^{(b)}_j\},
\end{equation}
where $\widehat{\lambda}_{h^{(b)}_j}(t)$ is the intensity function estimated using data from the terminal node $h^{(b)}_j$,
and the ensemble intensity, $\widehat{\lambda}^{(*)}(t;\bm{x})$, is computed by averaging over the $B$ trees:
\begin{equation} \label{eq:lambda-ensemble}
\widehat{\lambda}^{(*)}(t;\bm{x}) = \frac{1}{B}\sum_{b=1}^{B}\widehat{\lambda}^{(b)}(t;\bm{x})=   \frac{1}{B}\sum_{b=1}^{B}\sum_{j=1}^{M} \widehat{\lambda}_{h^{(b)}_j}(t)I\{\bm{x} \in h^{(b)}_j\}.
\end{equation}

Hence, when the dynamic covariates are incorporated, the intensity functions need to be estimated on a tree node instead of estimating MCF. Let $\tilde{h}$ denote either an intermediate or terminal node, we assume that the failure process on node $\tilde{h}$ can be modeled by a NHPP with the intensity function $\lambda_{\tilde{h}}(t)$ being parameterized by a vector $(\beta_0^{\tilde{h}},\beta_1^{\tilde{h}},...,\beta_q^{\tilde{h}})^T$ through a log-linear specification:
\begin{equation}
\log\lambda_{\tilde{h}}(t) = \beta_0^{(\tilde{h})} + \sum_{j=1}^{q}z_j(t) \beta_j^{(\tilde{h})}.
\end{equation}  

Then, $\lambda_{\tilde{h}}(t)$ can be estimated by minimizing the negative log-likelihood function with the $L^1$ penalty (Lasso): 
\begin{equation} \label{eq:likelihood}
  -\log L(\beta_0^{\tilde{h}},\bm{\beta}^{\tilde{h}}) + \omega
 ||\bm{\beta}^{\tilde{h}}||_1
\end{equation}  
where $\bm{\beta}^{\tilde{h}}=(\beta_1^{\tilde{h}},\beta_2^{\tilde{h}},...,\beta_q^{\tilde{h}})^T$ and $\log L(\beta_0^{\tilde{h}},\bm{\beta}^{\tilde{h}})= \sum_{i=1}^{n^{(\tilde{h})}}  ( -\int \lambda_{\tilde{h}}(s)ds + \sum_{j=1}^{k}\lambda_{\tilde{h}}(t_j) )$.
Here, $n^{(\tilde{h})}$ and $k$ are respectively the number of systems and pooled failure times on node $\tilde{h}$. The $L^1$ regularization shrinks some coefficients to be exactly zero. Although an increasingly larger number of covariates has been made available by the advances of sensing technologies, it is evident from our industry practice that many covariates are in fact redundant from either the statistical modeling or domain knowledge perspectives. Also note that, the likelihood (\ref{eq:likelihood}) can be evaluated in parallel for distributed datasets as the contribution to the total likelihood from each system can be calculated independently. 

The extended RF-R algorithm is summarized in Algorithm 3. The computational complexity of the extended RF-R algorithm depends on the dimension, $q$, of the dynamic covariates as well as the optimization methods used for maximizing the likelihood (\ref{eq:likelihood}). For example, the quasi-Newton method (the variable metric algorithm) is used in our numerical examples, and the computational complexity of the extended RF-R algorithm is $\mathcal{O}( m  n q^2 \log n)$, where $\mathcal{O}(q^2)$ is the cost associated with the quasi-Newton method.

The extended RF-R algorithm also satisfies the basic scalability, reliability and security requirements for big data analytics and no transfer of the raw data is required during the analysis; see Section \ref{sec:overview}. In Step 1, the data included in a bootstrap sample are generated locally on each \textit{worker} node in parallel. In Step 2, a tree is grown until the termination condition is met. At each node and for each splitting, the \textit{driver} node generates $m$ covariates at random from the $p$ covariates, and creates the candidate split point for each of the selected $m$ covariates. The \textit{worker} node performs the splitting of the local bootstrap data into two subpopulations, and computes the total likelihood for both subpopulations. The computed total likelihood, using data stored on the worker node, is returned to the \textit{driver} node. Then, the \textit{driver} node aggregates the local likelihood constructed from all worker nodes, and the optimum splitting covariate and split-point is carried out on the \textit{driver} node. The tree structure is updated and recorded on the \textit{driver} node until the termination condition is met. 

\singlespacing
\begin{algorithm}[H]
	
	\KwData{$\bm{y}_i$, $\bm{x}_i$ and $\bm{z}_i(t)$ for all $i$}
	
	\textbf{Step 1 }Draw $B$ bootstrap samples from the original data.
	
	\textbf{Step 2 }
	\For{b=1,...,B}{
		Grow a random tree by recursively repeating the following steps at each node of the tree:
		
		\textbf{2.1)} Select $m$ covariates at random from the $p$ covariates.
		
		
		\textbf{2.2)} Pick the best covariate and its split-point among the $m$ selected covariates by maximizing the $L^2$ distance of the intensity functions between the two daughter nodes; see (\ref{eq:inner_lambda}). At each daughter node, the intensity function, which depends on the dynamic covariates, is estimated by minimizing the negative log-likelihood function with the $L^1$ penalty (\ref{eq:likelihood})
		
		\textbf{2.3)} Grow the tree to full size under the condition that each terminal node contains at least $d_0$ systems with at least one failure. 
		
	}
	
	\textbf{Step 3 }Obtain the ensemble estimator by averaging all trees.
	
	\textbf{Step 4 }Compute the prediction error using the OOB data.
	\caption{The extended RF-R algorithm with static and dynamic covariates}
\end{algorithm}
\singlespacing

Based on the same idea described in Section \ref{sec:split}, the splitting rule for a node maximizes distance between two real-valued intensity functions in the $L^2$ space:
\begin{equation} \label{eq:inner_lambda}
|| \hat{\lambda}^{(-)}(t)-\hat{\lambda}^{(+)}(t)  ||_2 = \left( \int_{0}^{t_k}\left(\hat{\lambda}^{(-)}(t)-\hat{\lambda}^{(+)}(t)\right)^2dt \right)^{\frac{1}{2}} 
\end{equation}
where $\hat{\lambda}^{(-)}(t)$ and $\hat{\lambda}^{(+)}(t)$ respectively denote the estimated intensity functions on the two daughter nodes. 
The OOB ensemble intensity can also be calculated as:
\begin{equation} \label{eq:nhpp-oob}
\hat{\lambda}^{(\mathrm{OOB})}(t;\bm{x}) = \frac{\sum_{b=1}^{B}\gamma_i(b)\sum_{j=1}^{M} \hat{\lambda}_{h^{(b)}_j}(t)I\{\bm{x} \in h^{(b)}_j\}}{\sum_{b=1}^{B}\gamma_i(b)}
\end{equation}
where $\gamma_i(b)=1$ if system $i$ is not contained in the bootstrap sample $b$, otherwise, $\gamma_i(b)=0$. The OOB error can be measured using the C-index as described in Section \ref{sec:oob}.

\section{Numerical Examples} \label{sec:numerical}
Section \ref{sec:numerical} investigates two illustrative examples in order to generate some critical insights and demonstrate the key advantages of the proposed RF-R. The motivating example is re-visited in Section \ref{sec:case} to illustrate the application of RF-R on a real industrial problem. All data sets and computer code are available on Github (\url{https://github.com/dnncode/RF-R}).
\subsection{Numerical Example 1}
We start with a simple numerical example involving 200 systems. For each system, a number of 10 system attributes are respectively sampled from the unit interval $[0,1]$. Let $x_{i,j}$ represent the value of covariate $j$ associated with system $i$ ($i=1,2,...,200$, $j=1,2,...,10$), the failure data of system $i$ are simulated from a Homogeneous Poisson Process (HPP) with the following intensity: $\lambda_i=0.01$ if $\leq x_{i,1}, x_{i,2}\leq 0.5$, $\lambda_i=0.1$ if $0.5<x_{i,1}, x_{i,2}\leq 1$, otherwise $\lambda_i=0.05$. This data set is referred to as \textbf{\texttt{DATASET A}}. 
Note that, covariates $x_{3}, ..., x_{10}$ are purposely made redundant in this example. 
%

%
\begin{figure}%
	\centering
	\subfloat[]{{\includegraphics[width=.46\linewidth]{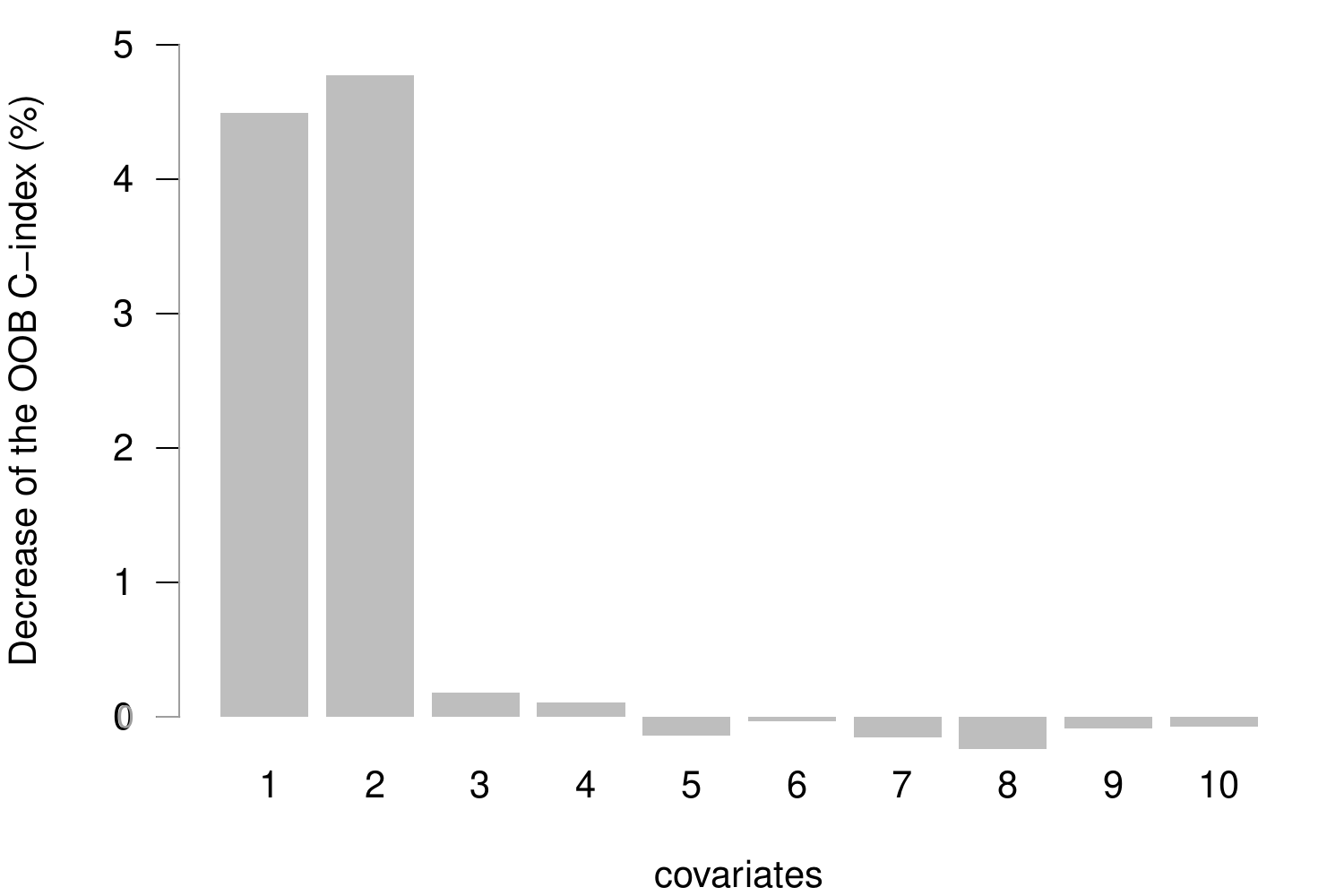} }}%
	\qquad
	\subfloat[]{{\includegraphics[width=.46\linewidth]{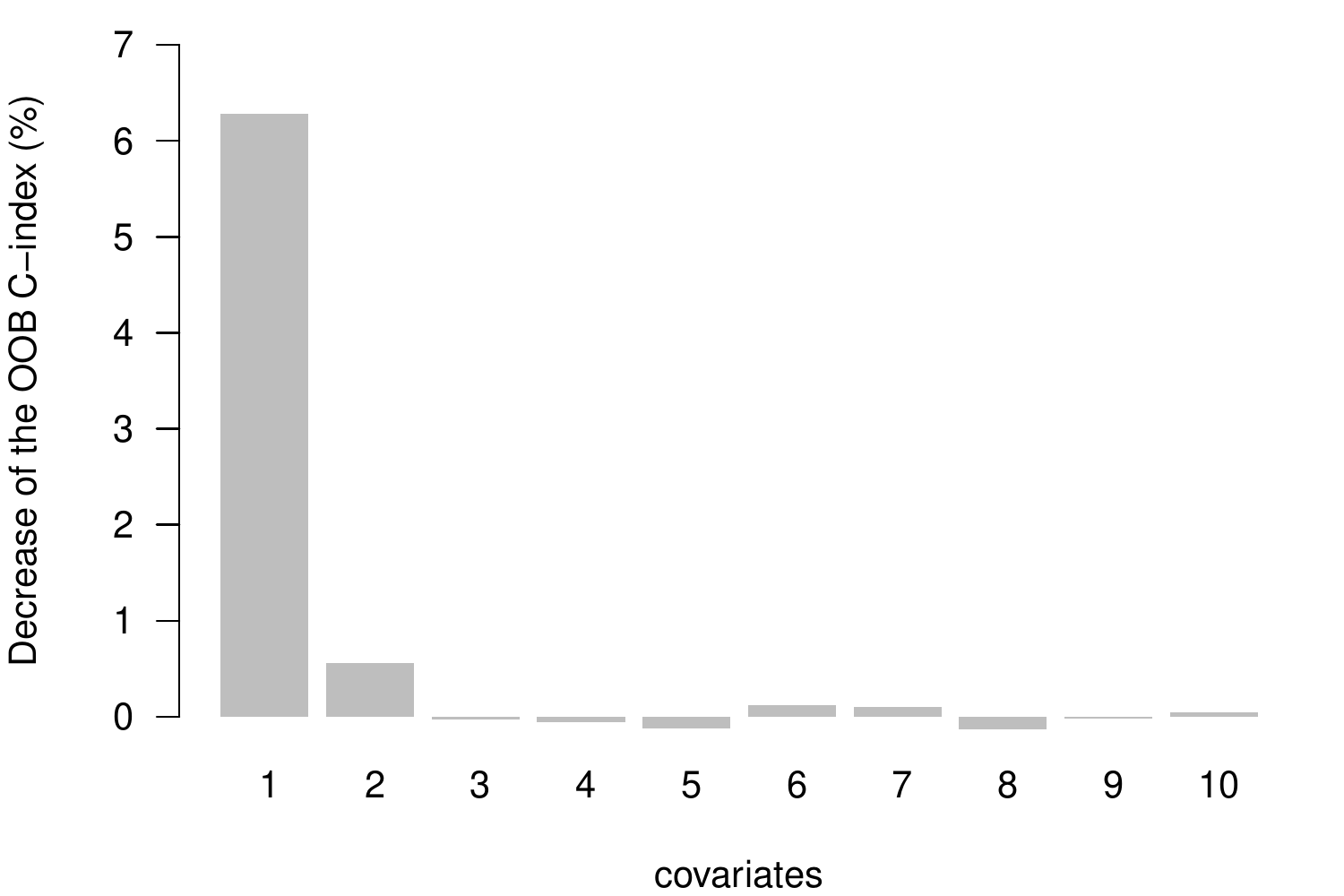} }}%
	\caption{Covariate importance based on \textbf{\texttt{DATASET A}} and \textbf{\texttt{DATASET B}}}%
	\label{fig:example1_vi} %
\end{figure}
Following the classical idea of identifying variable importance under the framework of RF, Figure \ref{fig:example1_vi}(a) shows the covariate importance as the average decrease of the OOB prediction C-index after the values of a particular covariate have been randomly permuted. A larger decrease of the OOB prediction C-index naturally indicates a stronger impact of a particular covariate on the prediction performance.
It is immediately seen that, the algorithm successfully identifies covariates $x_{1}$ and $x_{2}$ as important covariates, while covariates $x_{3},x_{4},...,x_{10}$ have little impact on the OOB prediction C-index and need to be excluded from the model.

\begin{figure}[h!]
	\begin{center}
		\includegraphics[width=1\textwidth]{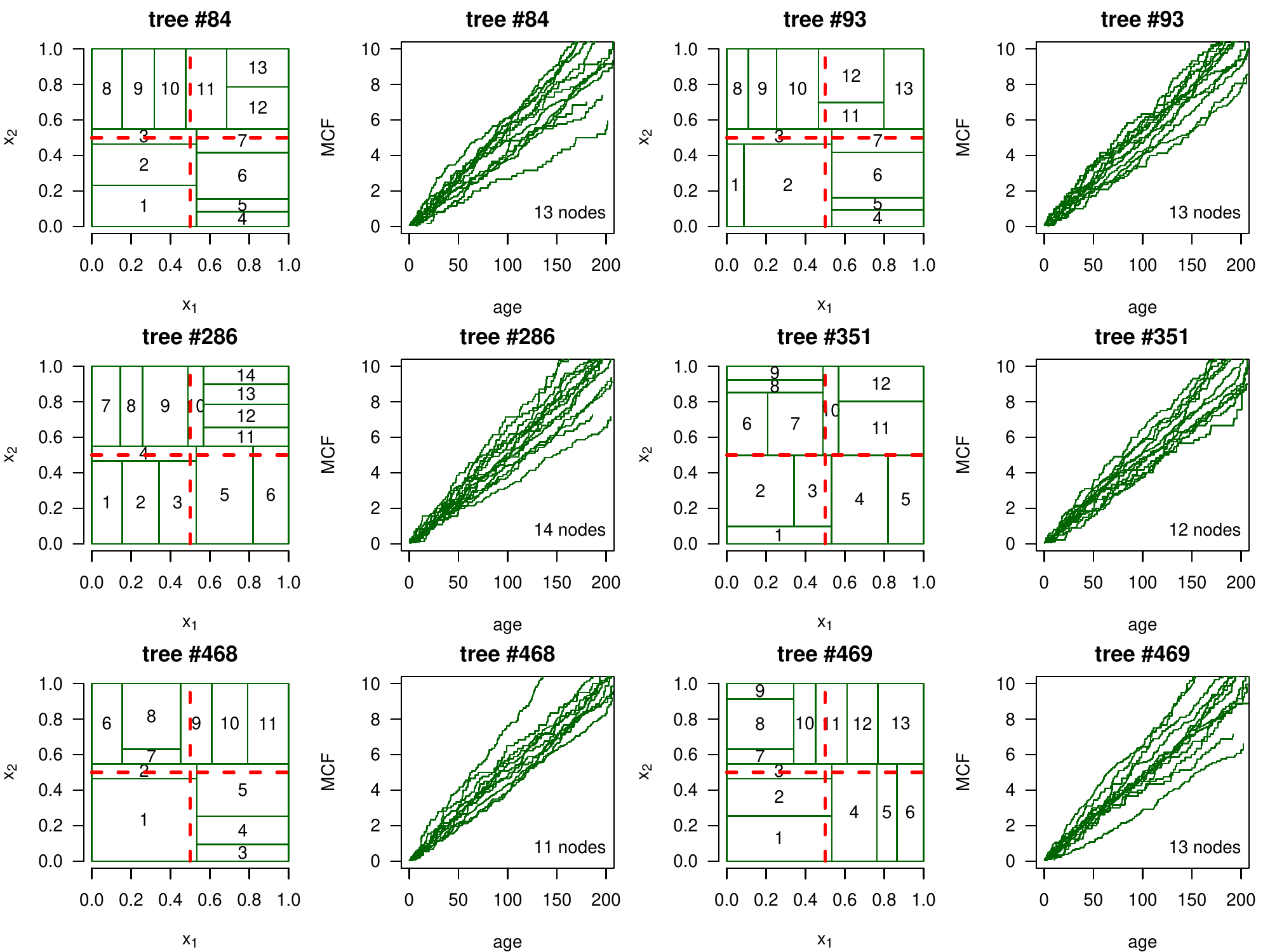}
	\end{center}
	\caption{Columns 1 and 3 show the binary partitions of the covariate space $[0,1]^2$ by eight randomly selected trees from the ensemble forests with the red dash lines indicating the true classification of the failure intensity. Columns 2 and 4 show the estimated MCF associated with each terminal node for the eight randomly chosen trees.}
	\label{fig:mcf_tree_example1} 
\end{figure}

We re-run the RF-R algorithm retaining only covariates $x_{1}$ and $x_{2}$. Since the space of the two covariates, $x_1$ and $x_2$, is a unit square $[0,1]^2$, it is possible to visualize the binary partition of the covariate space by each tree. Figure \ref{fig:mcf_tree_example1} shows the binary partition of the covariate space $[0,1]^2$ by eight randomly selected trees from the ensemble forests (with 500 trees). Note that, each tree divides the covariate space into a number of partitions, and each partition is represented by a terminal node. Hence, for each partition of a tree, the MCF is estimated using the data within that partition and is also shown in Figure \ref{fig:mcf_tree_example1}. 

For \textbf{\texttt{DATASET A}}, the intensity of the HPP falls into three classes depending on the values of $x_1$ and $x_2$. All trees in Figure \ref{fig:mcf_tree_example1} perform good binary partitions of the covariate space $[0,1]^2$. 
For example, tree \#84 partitions $[0,1]^2$ into 13 partitions (i.e., 13 terminal nodes). The first class, $x_{1}, x_{2} \in [0,0.5]$, is captured by terminal nodes 1 and 2; the second class, $x_{1}, x_{2} \in (0.5,1]$, is reasonably represented by terminal nodes $11\sim13$; while the third class is captured by the remaining terminal nodes. For another example, tree \#351 contains 12 terminal nodes. The first class, $x_{1}, x_{2} \in [0,0.5]$, is captured by terminal nodes $1\sim3$, and the second class, $x_{1}, x_{2} \in (0.5,1]$, is almost perfectly represented by terminal nodes $10\sim12$. 

Interestingly, Figure \ref{fig:mcf_tree_example1} may suggest the necessity to control the depth of the tree by combining some of the terminal nodes \citep{Huo2006}. For tree \#351, for example, the ideal case is to combine terminal nodes $1\sim3$ to form one bigger node. Note that, if the RF-R algorithm chooses the optimum splitting covariate and split point based on the log-rank test, one possible way to prune the tree is to stop further splitting a node if the minimum p-value is larger than a certain threshold, say, 0.05 or 0.1. This strategy effectively stops the tree from growing too deep. Alternatively, one might use a larger value of $d_0$ in the RF-R algorithm, which is the minimum number of systems with failures in a node. Fortunately, using full-grown trees seldom costs much in terms of model performance \citep{Hastie2009}. When trees are grown deep, each individual tree usually has low bias with high variance. Then, the idea of bagging (averaging full-grown trees) effectively reduces the variance of the ensemble estimator, and high model accuracy can thus be achieved. 

To demonstrate the advantages of the proposed approach, a cross-validation-based comparison study is performed considering the following four candidate approaches: 1) \textit{RF-R}: the proposed RF-R approach; 2) \textit{MCF}: the nonparametric estimation for MCF without utilizing any system attributes; 3) \textit{MCF-K}: the nonparametric estimation for MCF using only the data from the $K$ nearest ``neighboring'' a system. Here, the distance between two systems is given by the Euclidean distance based on their covariates; and 4) \textit{HPP}: the HPP model with a log-linear intensity, $\log(\lambda_i) = \beta_0 + \sum_{j=1}^{p}x_{i,j} \beta_j$ and $p=10$.
%
%
%
%
%
%

\begin{figure}[h!]
	\begin{center}
		\includegraphics[width=0.9\textwidth]{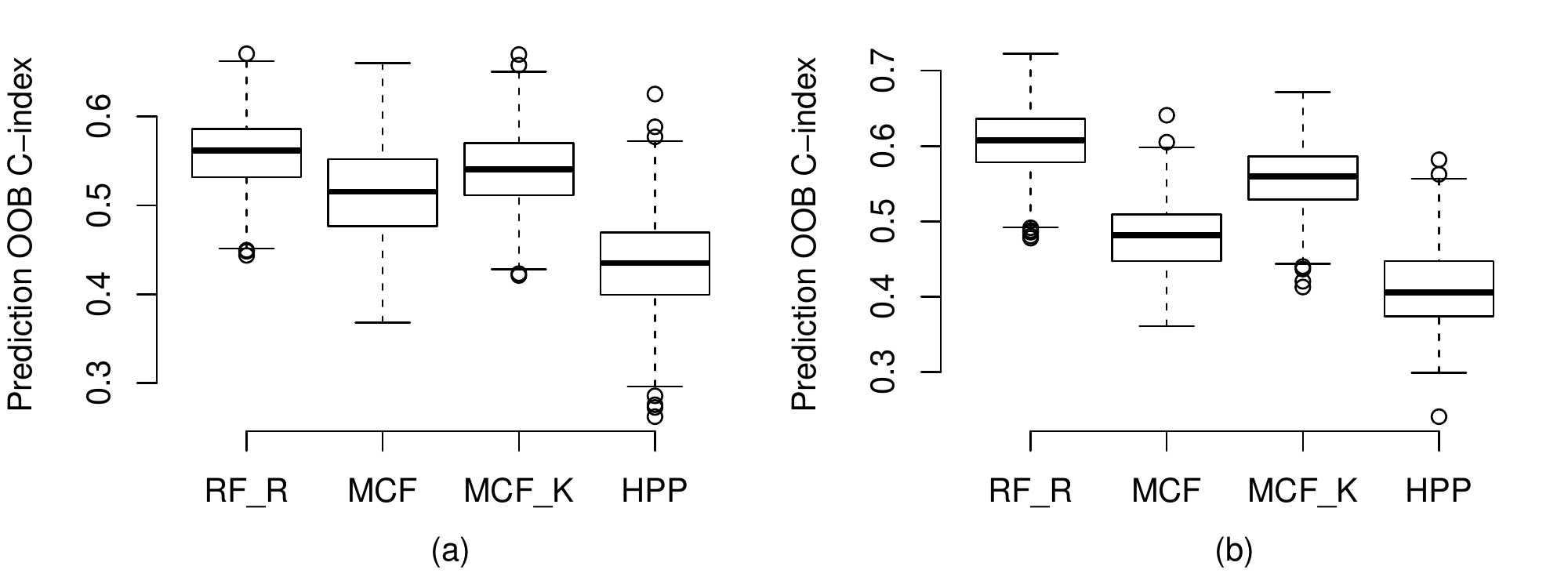}
	\end{center}
	\caption{Cross-validation-based comparison of the prediction C-index based on \texttt{\textbf{DATASET A}} and \texttt{\textbf{DATASET B}}.} 
	\label{fig:example1_compare} 
\end{figure}

Figure \ref{fig:example1_compare}(a) shows the cross-validation-based comparison of the prediction C-index for the four approaches above. A number of 500 iterations are performed. Within each iteration, the training and testing data sets are randomly split according to a 75\%-25\% ratio. We see that the proposed RF-R approach generates the highest prediction C-index. Note that, C-index is an evaluation metric based on pairwise ranking: 1) firstly, we rank the reliability of two systems based on the model; 2) then, we look at the observed reliability ranking of two systems, which is subject to uncertainty and may not always reflect the true reliability ranking of the two systems. In this sense, the C-index depends not only on the methods but also on the data itself. The comparison presented in Figure \ref{fig:example1_compare} is fair because different methods are applied to the same data set. 

One might argue that the assumption of HPP with a log-linear intensity function is not appropriate given how \texttt{\textbf{DATASET A}} is simulated. Hence, we regenerate the failure data exactly from a HPP with the log-linear intensity function $\log(\lambda_i) = \beta_0 +\beta_1  x_{i,1}  + \beta_2  x_{i,2}$,
where $\beta_0=0.01$, $\beta_1=2$ and $\beta_2=0.5$. This data set is referred to as \texttt{\textbf{DATASET B}}. Here, the covariates, $x_3,...,x_{10}$, are still treated as redundant variables. 
The comparison results based on \texttt{\textbf{DATASET B}} are shown in Figure \ref{fig:example1_compare}(b). We see that not only the RF-R method but also the two MCF-based methods outperform the parametric approach assuming HPP, even if the data are simulated exactly from a HPP. The performance of the parametric approach clearly suffers from the presence of eight redundant covariates. 

Based on \texttt{\textbf{DATASET B}}, Figure \ref{fig:example1_vi}(b) shows the covariate importance as the average decrease of the OOB prediction C-index. It is interesting to see that, since $\beta_1$ is set to be four times larger than $\beta_2$ when \texttt{\textbf{DATASET B}} is generated, the algorithm successfully identifies $x_1$ to be the most important covariate (all covariates are standardized). We also see that, although the importance of $x_2$ is much lower than that of $x_1$, $x_2$ is still obviously more important than the remaining eight redundant covariates, $x_3,..., x_{10}$, as shown in Figure \ref{fig:example1_vi}(b).
%
\subsection{Numerical Example 2}
We consider both static and dynamic covariates in the second numerical example. For each system, in addition to a number of 10 static covariates which are randomly sampled from the unit interval $[0,1]$, a dynamic time-varying covariate $z(t)$ is also simulated from a Brownian motion process  
$\sigma B_t$ where $B_t$ is a standard Brownian motion and $\sigma=0.1$. 
The failure data for 200 systems are generated. For any system $i$, its failure times are simulated from a NHPP using the thinning method \citep{Lewis1979} with the intensity function, $\lambda_i(t) = \exp\{\beta_{i,0}+\beta_{i,1} z_i(t)\}$, where $\exp\{\beta_{i,0}\}=0.01$ and $\beta_{i,1}=0.5$ if $0 \leq x_{i,1}, x_{i,2}\leq 0.5$, $\exp\{\beta_{i,0}\}=\beta_{i,1}=0.1$ if $0.5<x_{i,1}, x_{i,2}\leq 1$, and $\exp\{\beta_{i,0}\}=0.05$ and $\beta_{i,1}=0$ otherwise. 
This data set is referred to as \texttt{\textbf{DATASET C}}, and the failure process depends only on the first two attributes, $x_1$ and $x_2$, as well as the time-varying covariate $z(t)$. 

\begin{figure}[h!]
	\begin{center}
		\includegraphics[width=0.65\textwidth]{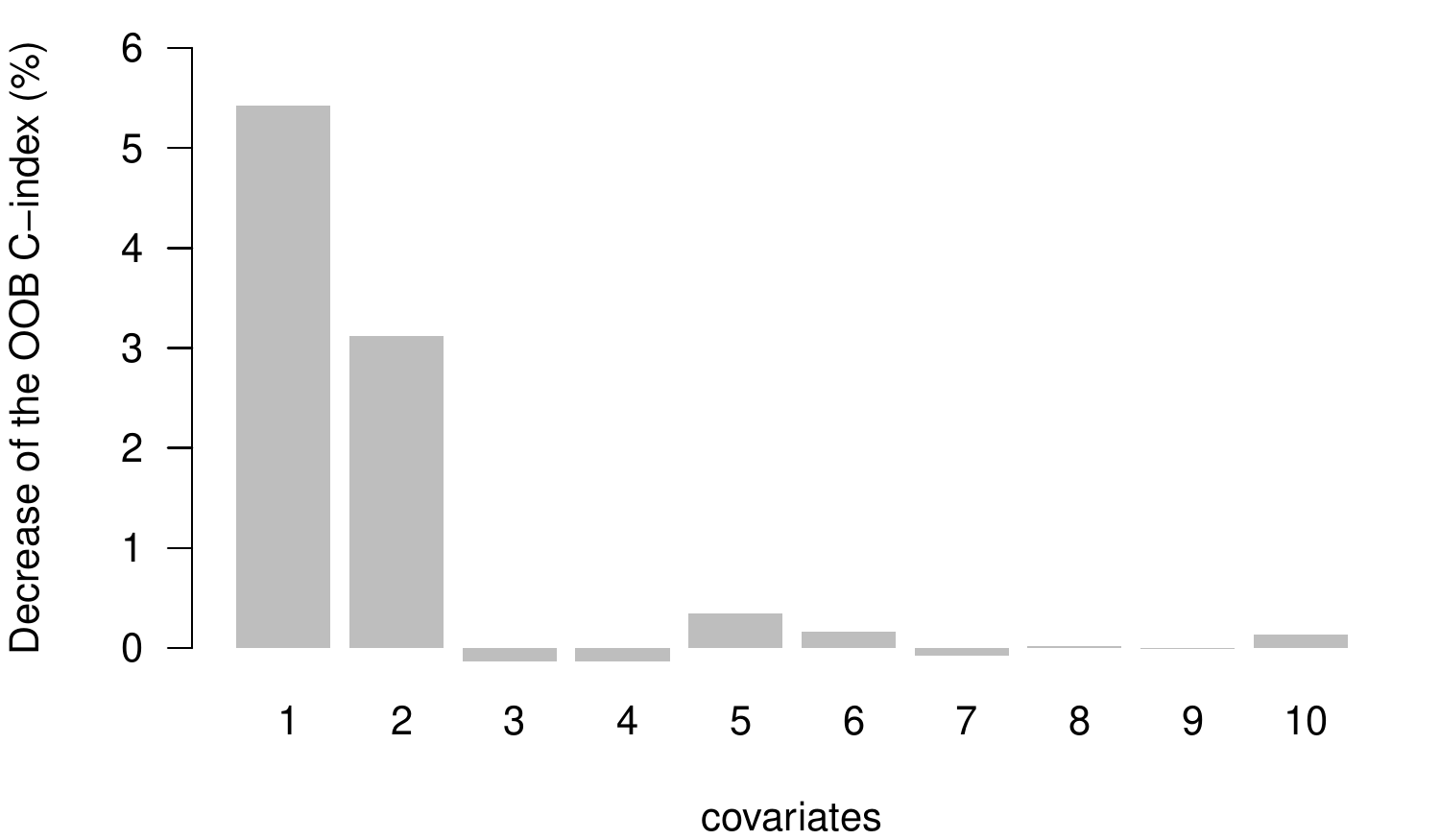}
	\end{center}
	\caption{Covariates importance based on \texttt{\textbf{DATASET C}}.}
	\label{fig:example2_vi_converge} 
\end{figure}

\begin{figure}[h!]
	\begin{center}
		\includegraphics[width=1\textwidth]{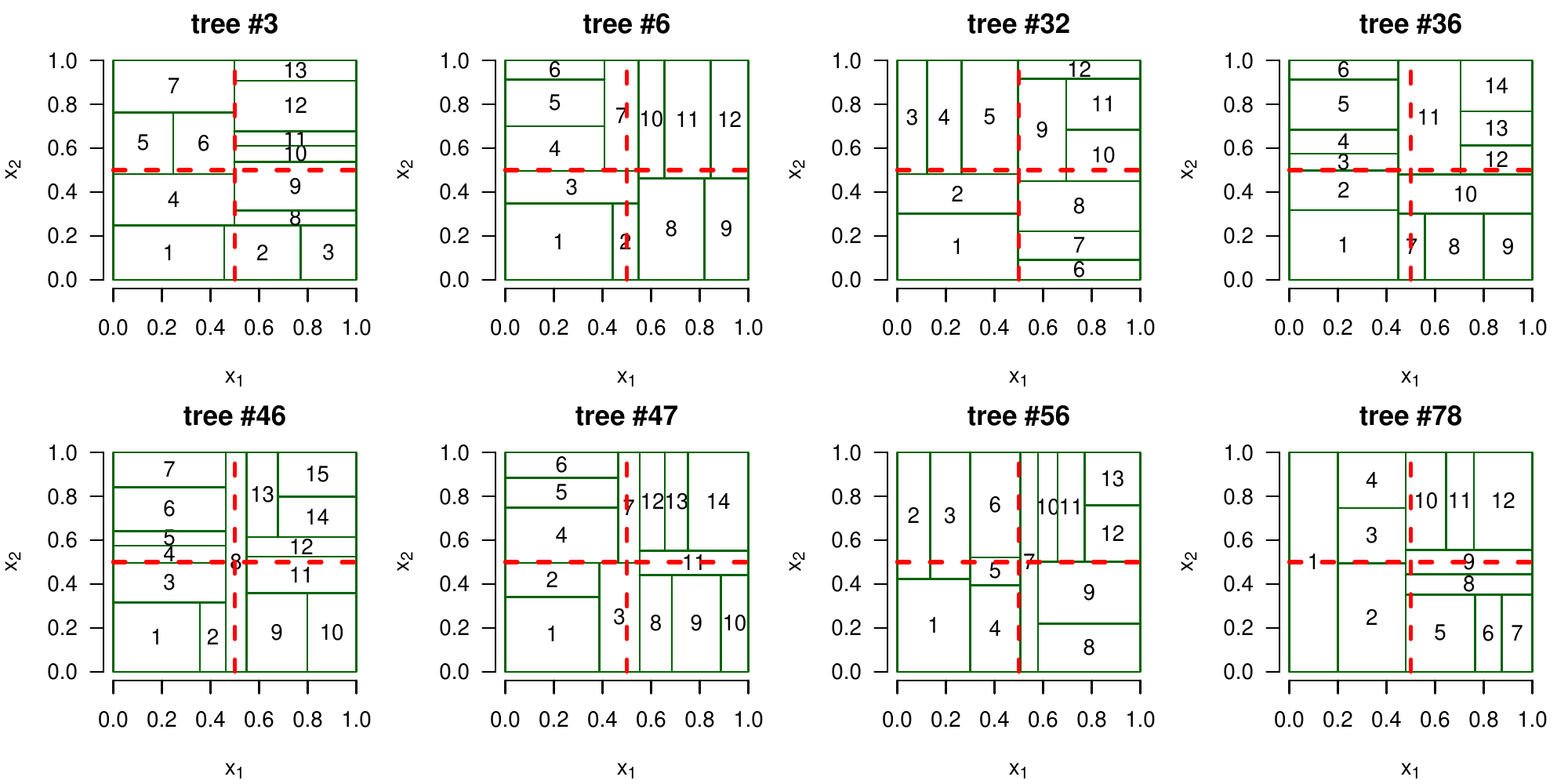}
	\end{center}
	\caption{Binary partitions of the covariate space by eight randomly chosen trees. The red dash lines indicate the true classification of failure intensity for \texttt{\textbf{DATASET C}}}
	\label{fig:mcf_tree_example2} 
\end{figure}

Figure \ref{fig:example2_vi_converge} shows the importance of the ten system attributes, $x_1$, $x_2$,...,$x_{10}$,  measured by the decrease of the OOB prediction C-index after the values of a particular attribute has been permuted. The algorithm successfully identifies the first two attributes, $x_1$ and $x_2$.
Similar to Figure \ref{fig:mcf_tree_example1} in Example 1, Figure \ref{fig:mcf_tree_example2} shows how the binary partitions of the space $[0,1]^2$ are performed by eight randomly chosen trees.  
For tree \#3, for example, the first class, $x_{1}, x_{2} \in [0,0.5]$, is captured by terminal nodes 1 and 4; the second class, $x_{1}, x_{2} \in (0.5,1]$, is reasonably well represented by terminal nodes $10\sim13$; while the third class is captured by remaining terminal nodes. Also note that, for an ensemble approach like RF, we generally do not expect all trees performs equally good, and some ``weak learners'' always exist.


\begin{figure}[h!]
	\begin{center}
		\includegraphics[width=0.9\textwidth]{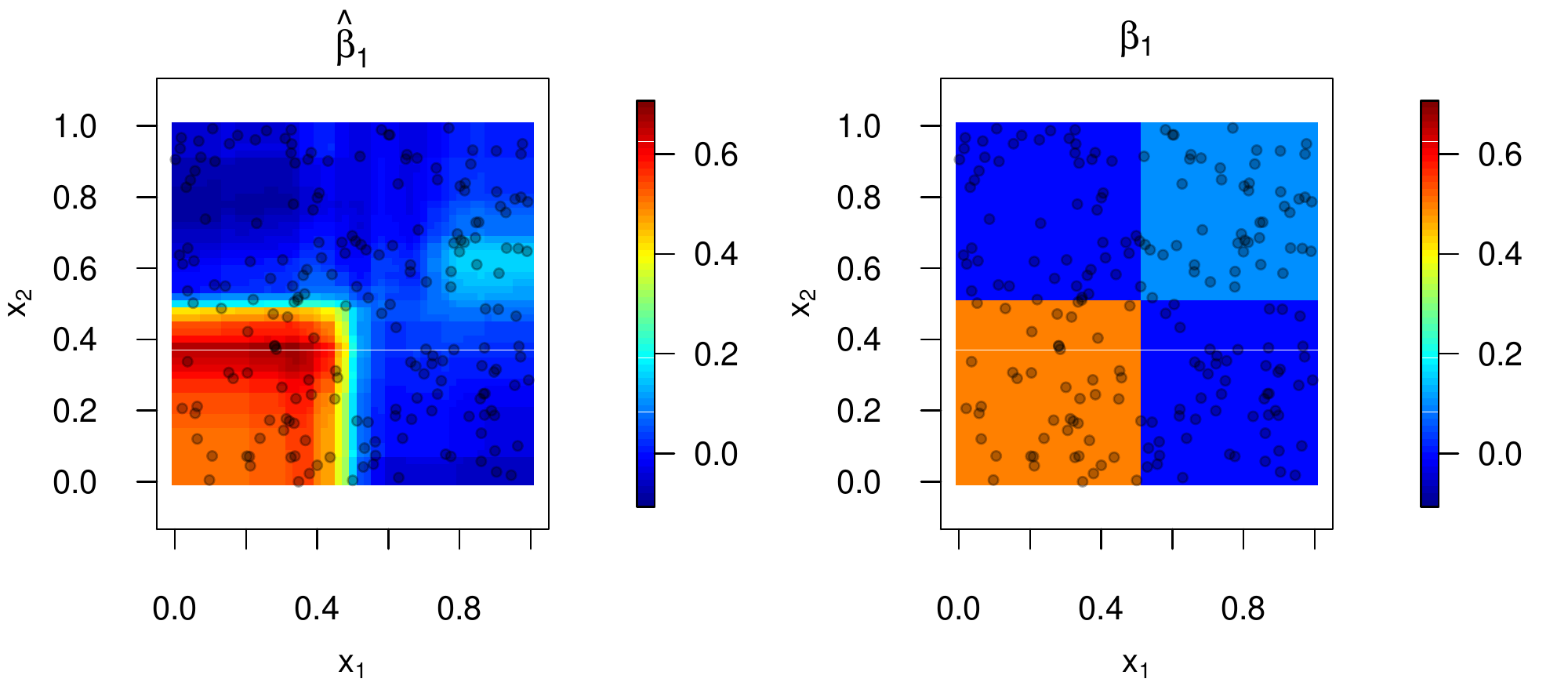}
	\end{center}
	\caption{Left panel: an aggregated view of the estimated effect $\hat{\beta}_1$ of the dynamic covariate for different $x_1$ and $x_2$; Right panel: the true effect ${\beta}_1$ for \texttt{\textbf{DATASET C}}.}
	\label{fig:example2_beta_spatial} 
\end{figure}

More interestingly, to show how the RF-R effectively captures the interactions between the static covariates, $x_1$ and $x_2$, and the dynamic covariate, $z(t)$, Figure \ref{fig:example2_beta_spatial} provides an aggregated view of the effect $\hat{\beta}_1$ from the ensemble trees over the domain $[0,1]^2$ (note that, each tree partitions the space $[0,1]^2$ into a number of rectangular areas and the effects ${\beta}_1$ are estimated for each area).
In this figure, the panel on the left shows how the estimated effect $\hat{\beta}_1$ interacts with the static covariates, while the panel on the right shows the actual effect of $\beta_1$ (recall that the true effect ${\beta}_1$ of the dynamic covariate is much stronger when $x_1$ and $x_2$ are below 0.5 for \texttt{\textbf{DATASET C}}). We see from this figure that the proposed RF-R algorithm successfully captures the interaction between the dynamic and static covariates, which can be extremely challenging for conventional approach. 

\begin{figure}[h!]
	\begin{center}
		\includegraphics[width=1\textwidth]{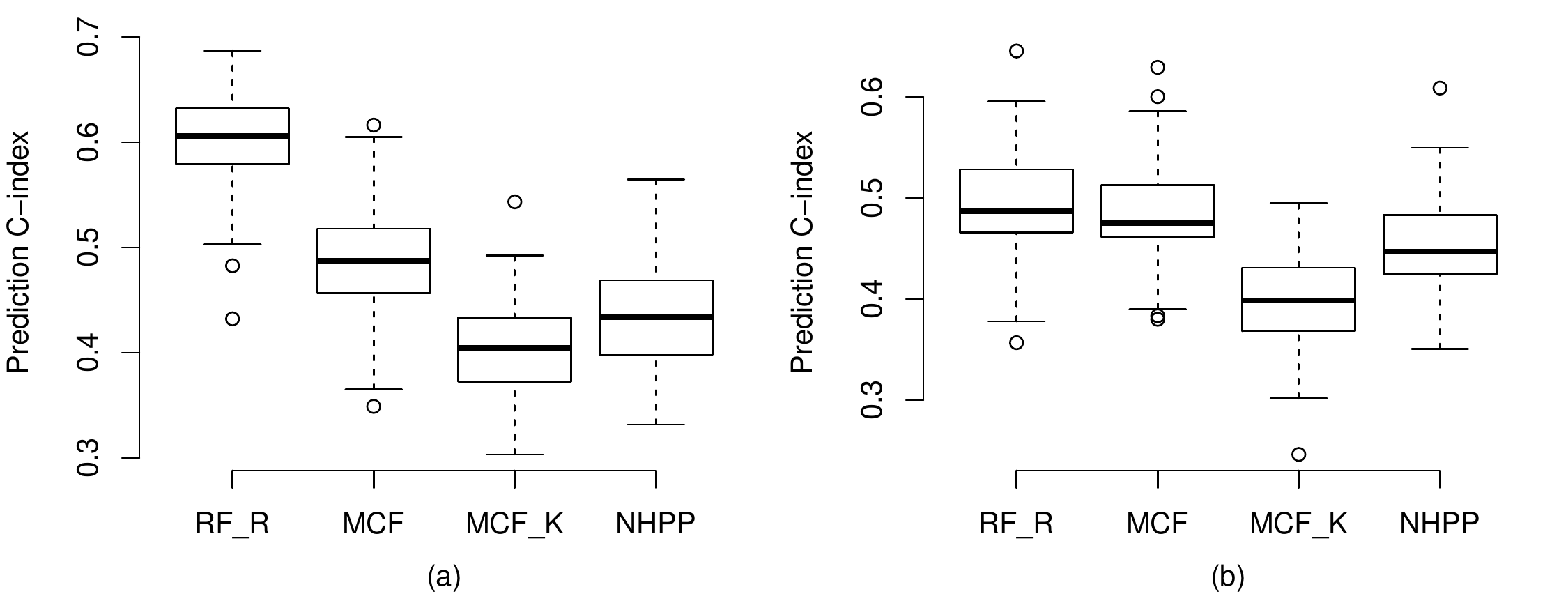}
	\end{center}
	\caption{Cross-validation-based comparison of the prediction C-index for four different approaches. Panel (a) on the left shows the comparison results based on \texttt{\textbf{Dataset C}}, and panel (b) on the right shows the results based on \texttt{\textbf{Dataset D}}.}
	\label{fig:example2_comparison} 
\end{figure}

A cross-validation-based comparison of the C-index is performed for RF-R, MCF, MCF-K and NHPP which refers to the MLE assuming a NHPP model with a log-linear intensity function, $\lambda_i(t) = \exp\{\beta_{i,0}+\sum_{j=1}^{10}\beta_{i,j}x_{i,j} + \beta_{i,11} z_i(t)\}$.
Figure \ref{fig:example2_comparison}(a) shows the comparison based on \texttt{\textbf{DATASET C}}. We see that the RF-R again outperforms in terms of the prediction C-index. Note that, since the NHPP model is not appropriate given how \texttt{\textbf{DATASET C}} is simulated, we re-generate the data using the intensity function, $\lambda_i(t) = \exp\left\{\beta_{0}+\beta_{1}x_{i,1} +\beta_{2}x_{i,2}+ \beta_{3} z_i(t)\right\}$
where $\beta_{0}=\log(0.01)$, $\beta_1=2$, and $\beta_2=\beta_3=0.5$. This dataset is referred to as \texttt{\textbf{Dataset D}}. Based on \texttt{\textbf{Dataset D}}, we regenerate the comparison results as shown in Figure \ref{fig:example2_comparison}(b). The proposed RF-R still outperforms. This figure also indicates that the presence of redundant covariates $x_3 \sim x_{10}$ has a significant detrimental impact on the prediction accuracy of the NHPP model. However, the proposed RF-R algorithm appears to be much more robust against redundant covariates, which is a salient advantage of the proposed method in modern Big Data environment where the presence of redundant covariates is almost inevitable. 

\section{Case Study: The Motivating Example Revisited} \label{sec:case}
The motivating example in Section \ref{sec:motivating} is re-visited in this section to demonstrate the application of the proposed approach on a real problem arising from industry. 
This case study is based on a modified data set which consists of 8232 oil and gas wells installed over 2007$\sim$2017. More details can be found in Section \ref{sec:motivating} and the data is available from the GitHub (\url{https://github.com/dnncode/RF-R}).

\begin{figure}[h!]
	\begin{center}
		\includegraphics[width=1\textwidth]{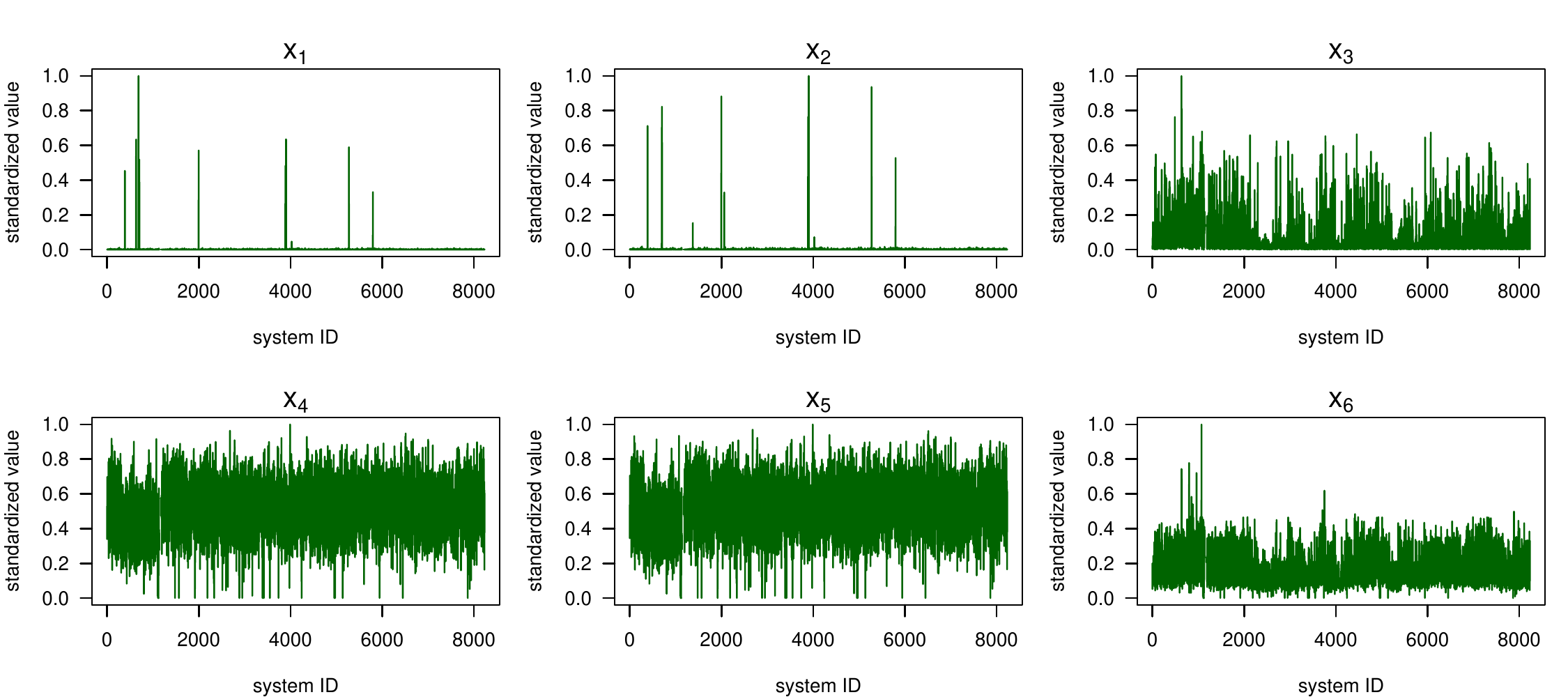}
	\end{center}
	\caption{Standardized values of $x_1 \sim x_{6}$ for all 8232 systems.}
	\label{fig:X_case} 
\end{figure}

A number of eight well attributes $x_1 \sim x_8$ are available and all covariates are standardized on the unit interval $[0,1]$. In particular, the geo-locations of these wells are given by covariates $x_7$ (latitude) and $x_8$ (longitude); see Figure \ref{fig:wells}. Covariates $x_1 \sim x_6$ are static attributes that contain basic well characteristics. Figure \ref{fig:X_case} shows the standardized values of the six attributes for all 8232 systems, and the heterogeneity among systems is clearly shown.

\begin{figure}[h!]
	\begin{center}
		\includegraphics[width=1\textwidth]{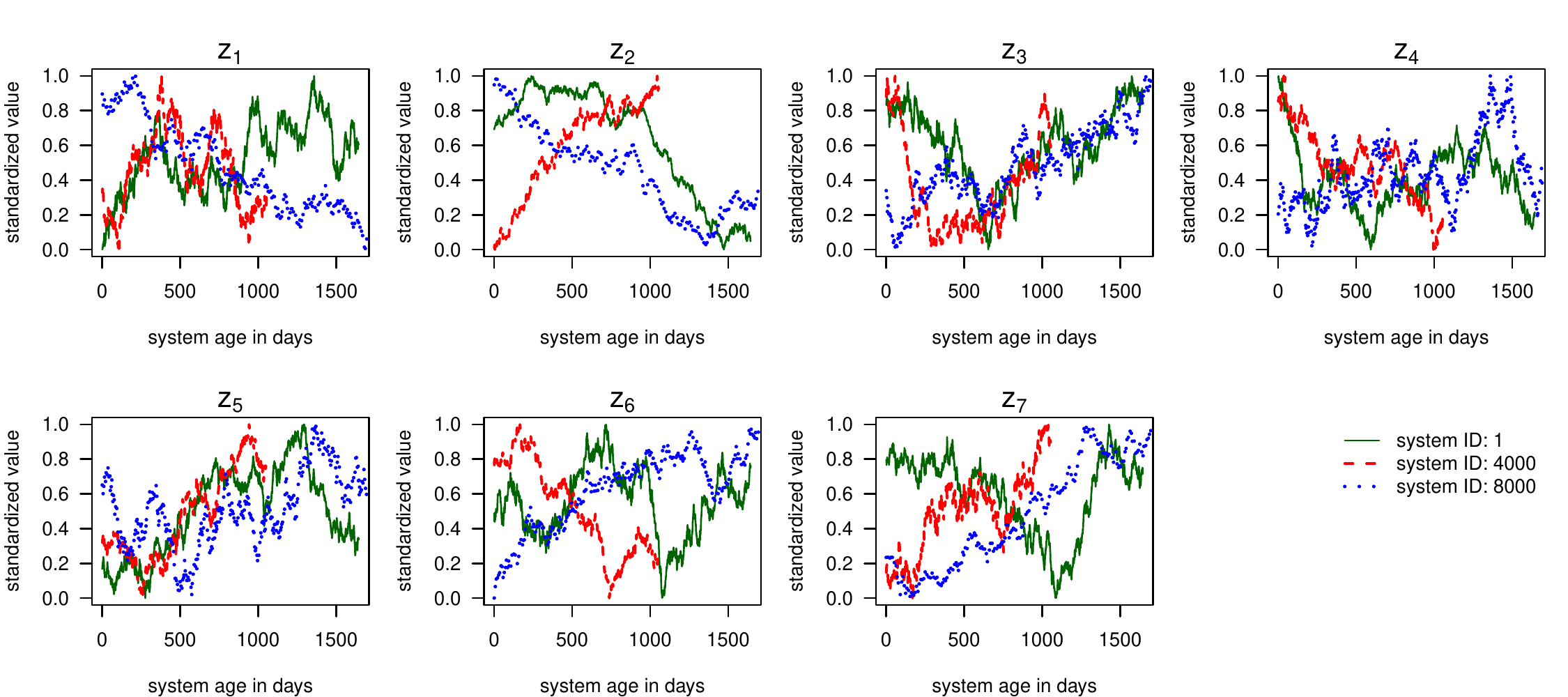}
	\end{center}
	\caption{Sensor measurement in terms of torque, load and stress, $z_1 \sim z_7$, for three selected systems: system \#1, system \#4000, and system \#8000.}
	\label{fig:Z_case} 
\end{figure}

In addition to the static covariates $x_1 \sim x_8$, seven dynamic sensor measurement of system operating conditions, $z_1(t) \sim z_7(t)$, are also available. In particular, $z_1(t) \sim z_5(t)$ are related to the dynamic torque and load measurement of each well including gear torque, load range, and so on, while $z_{6}(t)$ and $z_{7}(t)$ are respectively the monitored gearbox and structural stress. The sensor data are aggregated on daily basis. For illustrative purposes, Figure \ref{fig:Z_case} plots the values of $z_1(t) \sim z_7(t)$ for three systems: system \#1, \#4000, and \#8000. We see that, different well systems experience very different operating conditions in terms of torque, load and stress, further increasing the heterogeneity among these systems.

\begin{figure}[h!]
	\begin{center}
		\includegraphics[width=0.9\textwidth]{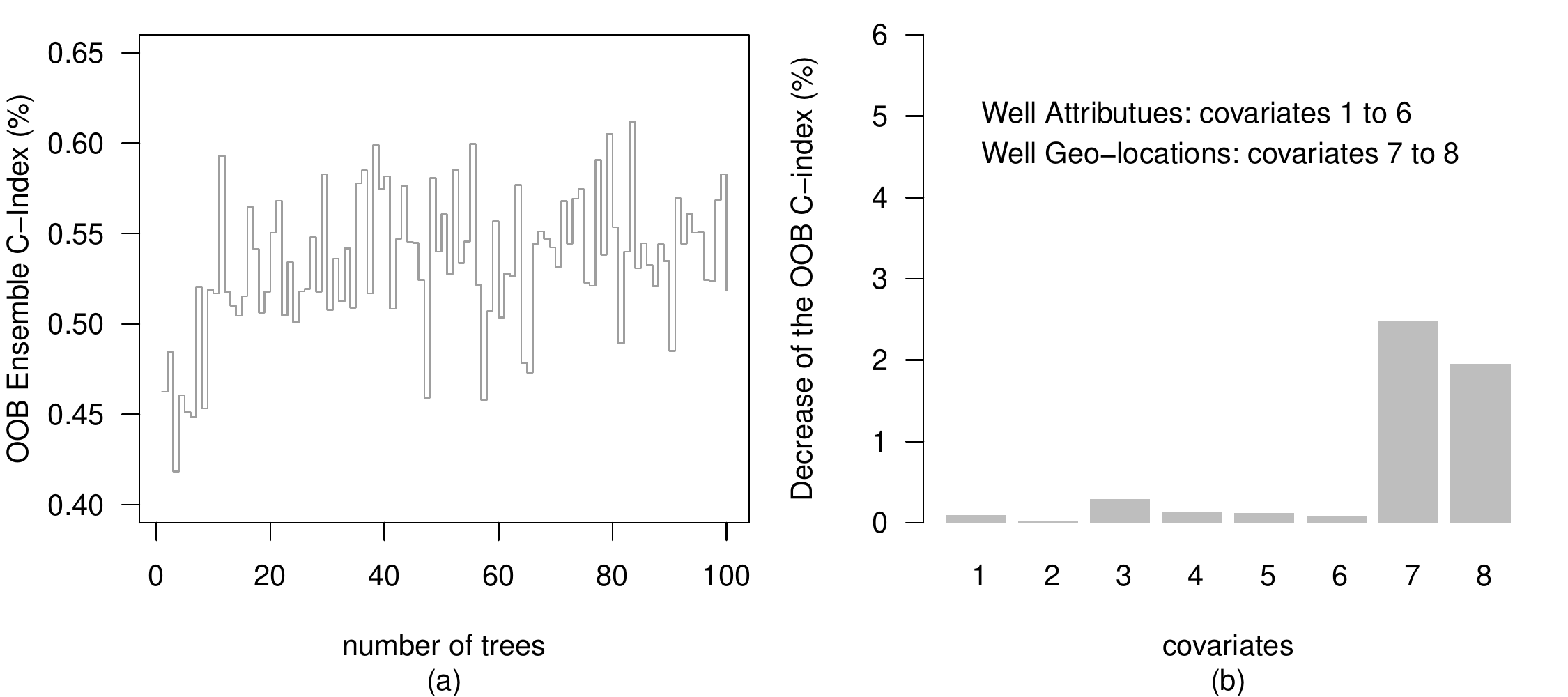}
	\end{center}
	\caption{Panel (a) shows the OOB ensemble prediction C-index against the number of trees in a forest for the first 100 trees, and panel (b) shows the importance of the eight static system attributes measured by the average decrease of the OOB prediction C-index}
	\label{fig:case_vi} 
\end{figure}

We run the RF-R algorithm, and Figure \ref{fig:case_vi}(a) shows the OOB ensemble prediction C-index against the number of trees in a forest for the first 100 trees. It is seen that, the OOB prediction C-index quickly increases as the number of trees increases, and is stabilized approximated after 50 trees have been grown. 
Figure \ref{fig:case_vi}(b) shows the importance of the eight static system attributes as the average decrease of the OOB prediction C-index after the values of a particular covariate has been randomly permuted. Interestingly, the algorithm identifies $x_7$ and $x_8$, the geo-locations, as the two most important system attributes. 
Because wells at similar geographical locations share common, but unknown, environmental conditions (e.g., temperature and humidity variation, soil type, contamination, etc), geo-locations may serve as important proxies in capturing those unknown environmental factors which may lead to some important spatial patterns such as trend and clustering. 
The results shown in Figure \ref{fig:case_vi}(b) confirm that these unknown environmental factors significantly influence the system failure processes, causing distinctive failure patterns among well systems. One might also note that the importance of latitude, $x_7$, is slightly higher than that of longitude, $x_8$, indicating a stronger spatial trend along the south-north direction. This is consistent with our initial observations from Figures \ref{fig:wells}.

\begin{figure}[h!]
	\begin{center}
		\includegraphics[width=0.9\textwidth]{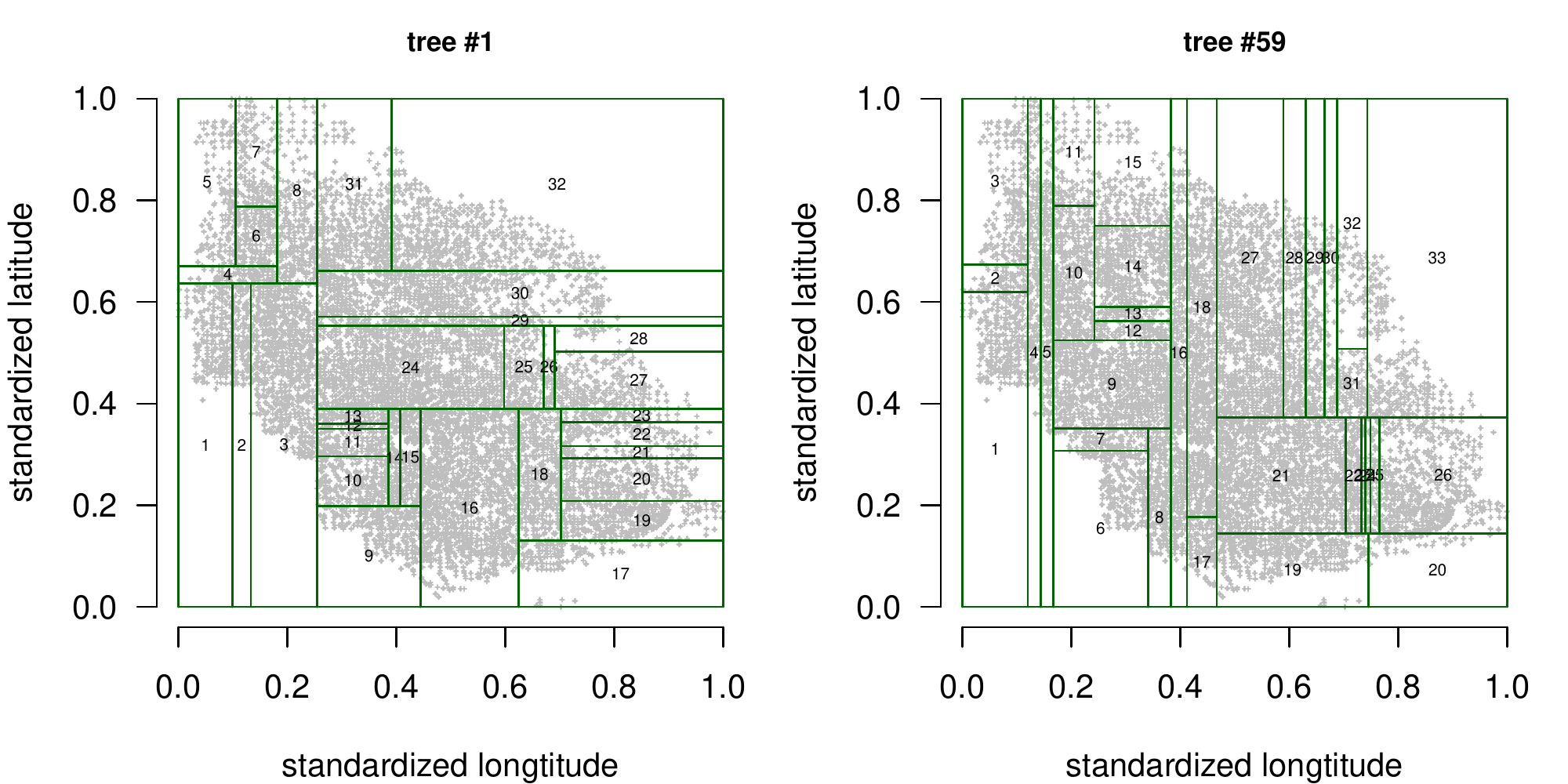}
	\end{center}
	\caption{Binary partition of the spatial domain by two chosen trees: tree \#1 and \#59.}
	\label{fig:casestudy1_partition} 
\end{figure}

\begin{figure}[h!]
	\begin{center}
		\includegraphics[width=1\textwidth]{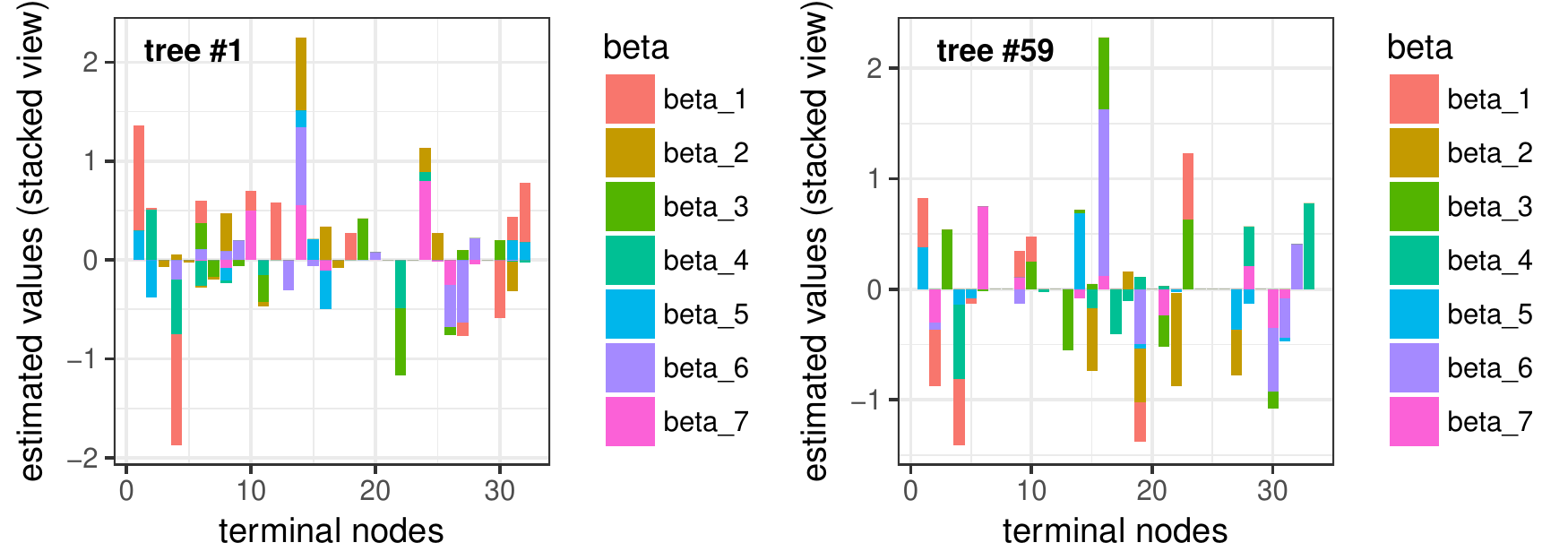}
	\end{center}
	\caption{The estimated values, $\hat{\beta}_1 \sim \hat{\beta}_7$, on each terminal node of tree \#1 and tree \#59.}
	\label{fig:casestudy1_lasso} 
\end{figure}

To generate more insights on how the RF-R performs, we re-run the analysis by only retaining the geo-location information, $x_7$ and $x_8$, and the sensor measurement, $z_1(t) \sim z_7(t)$. For illustrative purposes, Figure \ref{fig:casestudy1_partition} shows the binary partition of the spatial domain from two chosen trees: tree \#1 and \#59. Tree \#1 divides the spatial domain into 32 rectangular regions, while tree \#59 divides the spatial domain into 33 rectangular regions. At each terminal node of a tree, the intensity function of the NHPP, which depends on the dynamic conditions measured by sensors, is estimated by minimizing the negative log-likelihood with the $L^1$ penalty in (\ref{eq:likelihood}). Note that, not all sensor measurement are relevant in terms of estimating the failure process. 
As an illustrative example, Figure \ref{fig:casestudy1_lasso} shows the estimated values, $\hat{\beta}_1 \sim \hat{\beta}_7$, on every terminal node of tree \#1 and tree \#59. This figure is presented in the stacked view in the sense that, for each terminal node, the heights of the colored bars respectively indicate the sizes of the estimated values. It is seen that, the Lasso penalty successfully imposes sparsity on the terminal nodes of a tree: at each terminal node, only a small subset of $z_1(t) \sim z_7(t)$ is included in the estimated intensity function, which describes the failure process of the systems on that terminal node. This result also suggests that the intensity functions of systems from different geo-regions may have different dependence structures on system operational and environmental conditions, indicating a high level of heterogeneity among the 8232 systems in this case study. 

\begin{figure}[h!]
	\begin{center}
		\includegraphics[width=1\textwidth]{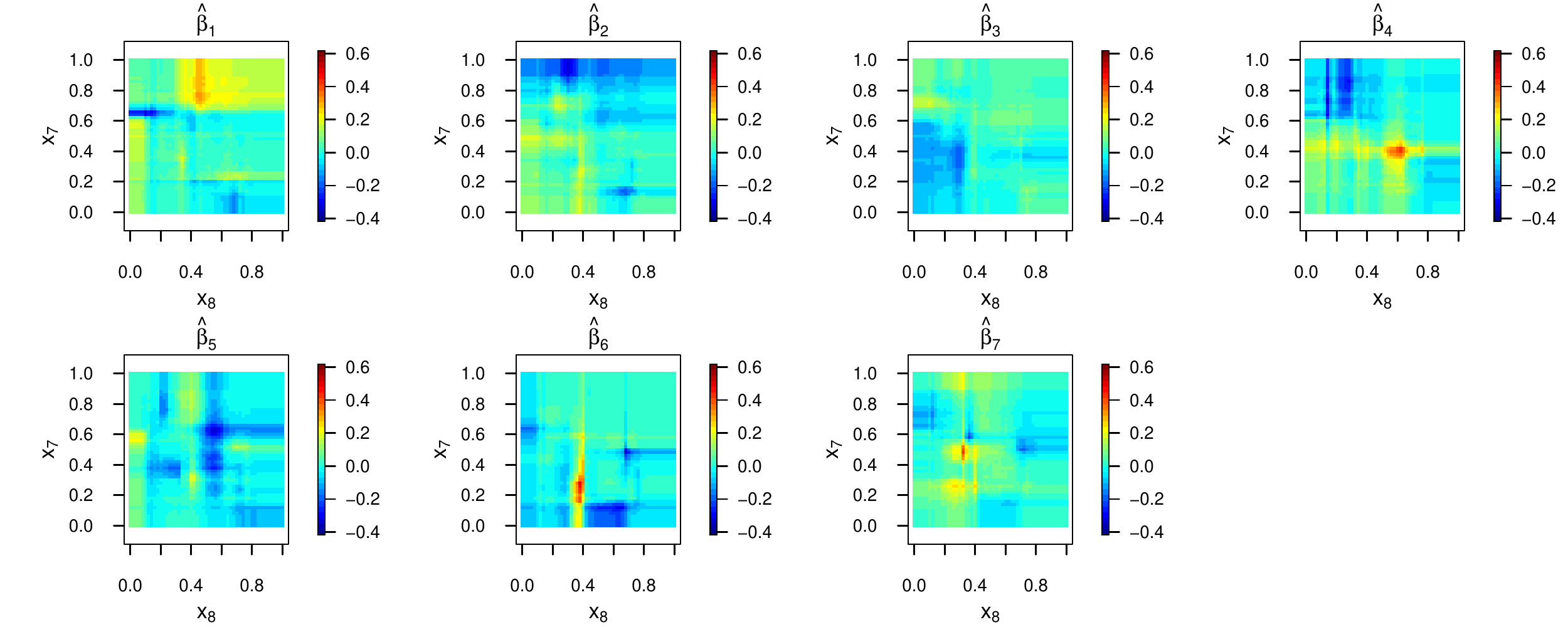}
	\end{center}
	\caption{A spatially aggregated view of the averaged $\hat{\beta}_1 \sim \hat{\beta}_7$ over the spatial domain.}
	\label{fig:casestudy1_beta_spatial.pdf} 
\end{figure}

To further elaborate the interactions between geo-locations and system operating conditions, Figure \ref{fig:casestudy1_beta_spatial.pdf} provides a spatially aggregated view of the averaged $\hat{\beta}_1 \sim \hat{\beta}_7$ from the ensemble trees over the spatial domain. Note that, each tree partitions the spatial domain $[0,1]^2$ into a number of rectangular areas and the estimates $\hat{\beta}_1 \sim \hat{\beta}_7$ are obtained for each area.  It is immediately seen that the failure intensities at different geo-locations are dominated by different operational and environmental conditions. For example, $z_1$ appears to have a positive effect on the intensity function for systems located in the area where $x_7>0.7$ and $x_8>0.4$ (i.e., the top right area of the first subplot on the first row), $z_2$ generally has a negative effect on the intensity function for systems located to the area where $x_7>0.9$ (i.e., the top area of the second subplot on the first row), and so on. Some sensor measurements may have particularly strong local effect on the failure intensity. For example, $z_4$ has a strong effect on the intensity function for those systems in the area where $x_7$ and $x_8$ are respectively close to 0.4 and 0.6 (i.e., the central area of the first subplot on the second row), and $z_6$ has a particularly strong effect on the intensity function for the systems in the area where $x_7$ and $x_8$ are respectively close to 0.2 and 0.4, and so on. These observations yield critical insights on system reliability and can be potentially useful for system maintenance and operation. 

\begin{figure}[h!]
	\begin{center}
		\includegraphics[width=0.65\textwidth]{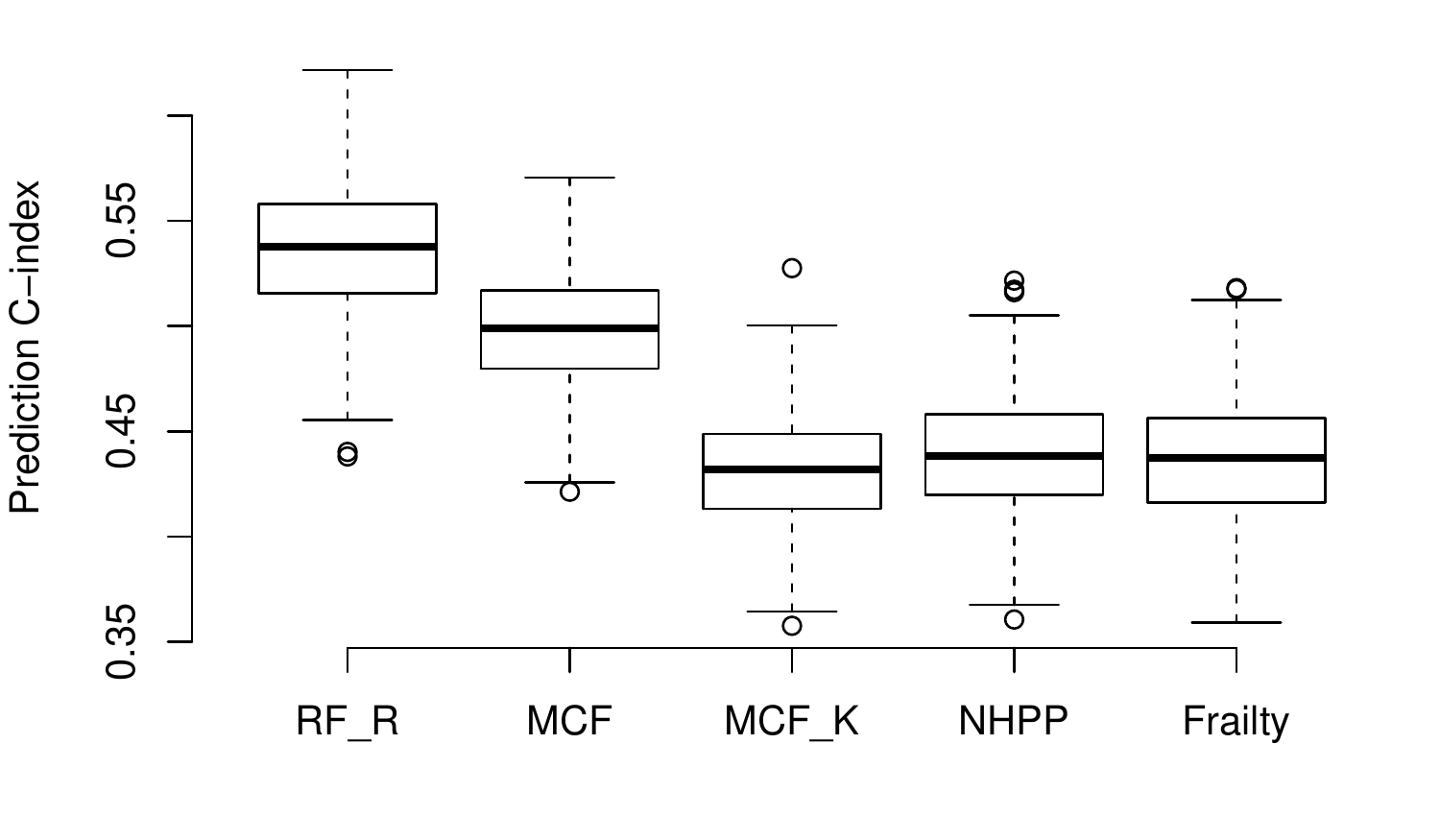}
	\end{center}
	\caption{Cross-validation-based comparison of the prediction C-index for four different approaches for the case study}
	\label{fig:casestudy1_comparison} 
\end{figure}

We compare the C-index, using cross-validation, between the RF-R, MCF, MCF-K, NHPP with a log-linear intensity function, and the Gamma frailty model. The Gamma frailty model is a commonly used approach to capture the random variation among systems; see \cite{Lindqvist2006}. Figure \ref{fig:casestudy1_comparison} shows the comparison result. The proposed RF-R again outperforms in terms of the prediction C-index. The parametric approaches, such as the NHPP and Gamma frailty model, suffer from the presence of a large number of redundant covariates. The fact that MCF has the second best performance suggests that we would rather not to use any covariate information than incorporating a large number of redundant covariates. 

\begin{figure}[h!]
	\begin{center}
		\includegraphics[width=0.9\textwidth]{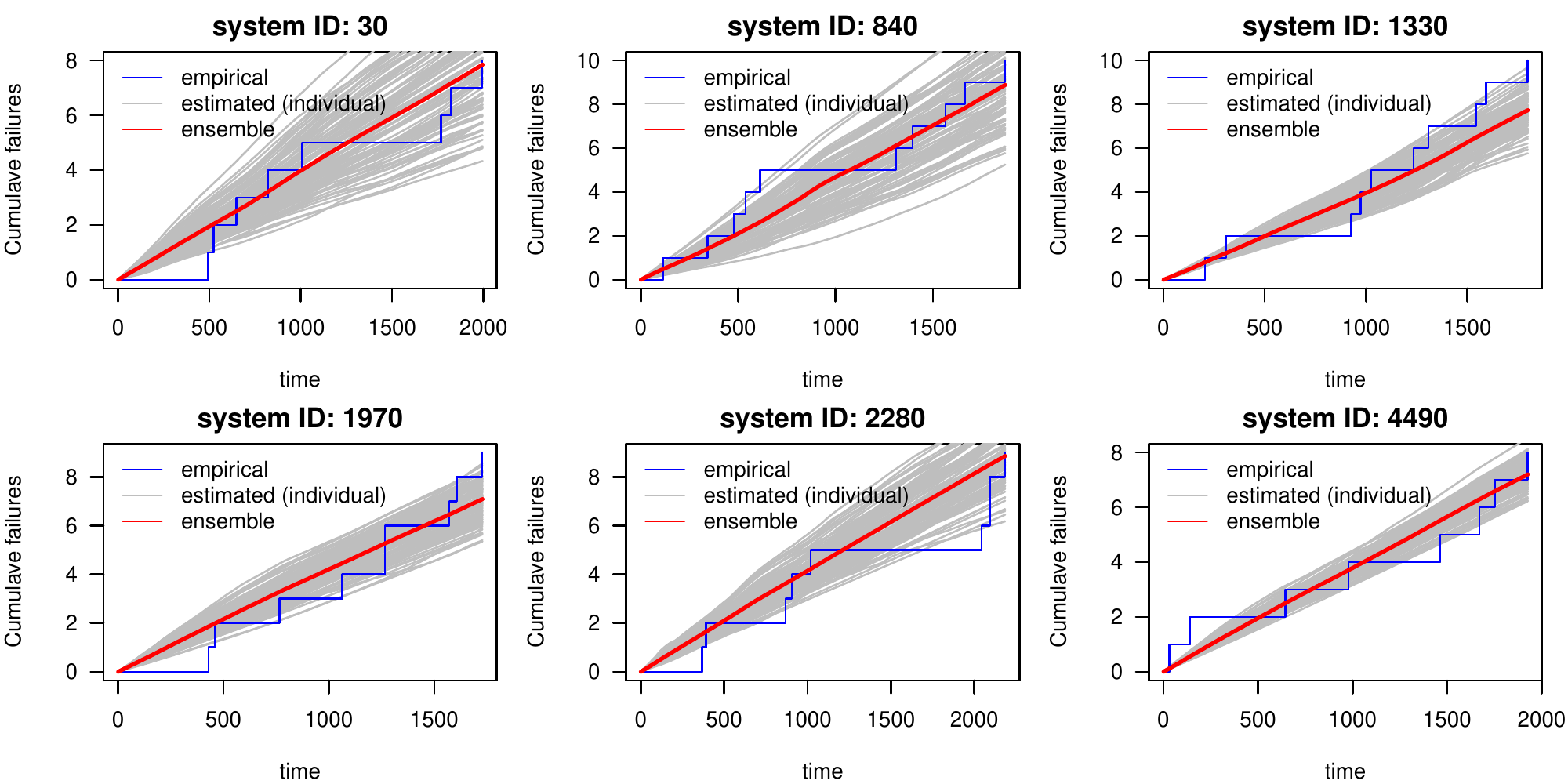}
	\end{center}
	\caption{Estimated cumulative hazard for selected systems.}
	\label{fig:casestudy1_mcf_individual} 
\end{figure}

In repairable system reliability analysis, the cumulative hazard function is often of interest \citep{Meeker1998a}. Hence, a number of 6 systems are selected from the 8232 wells. Figure \ref{fig:casestudy1_mcf_individual} shows both the true cumulative failures over time and the estimated cumulative hazard functions for the selected systems. To capture the uncertainty, this figure also includes the estimated cumulative hazard from individual trees. We see that, the RF-R effectively models the system reliability based on the recurrence data from a large fleet of repairable systems with both static and dynamic covariates.

\section{Conclusions} \label{sec:conclusions}
An algorithm called RF-R has been proposed for analyzing large heterogeneous repairable system reliability data with static system attributes and dynamic sensor measurement. Comprehensive numerical studies and comparison have been performed and the advantages of the proposed method have been demonstrated. The strengths of the proposed algorithm lie in the integration of the powerful Random Forests algorithm and the classical statistical reliability data analysis methodologies. This work timely addressed some pressing challenges facing reliability engineers today, including system heterogeneity, covariate selection, model specification and data locality in the modern Big Data environment. Computer program and data sets are made available on GitHub to facilitate the adoption and application of the proposed method in practice. 

\appendix
\section*{Appendix: Proof of Propositions 1 and 2}


Proposition 1 shows the uniform consistency of the ensemble RF-R estimator. To prove this proposition, note that,
\begin{equation} \label{eq:consistencyproof1}
\begin{split}
\lim_{N\rightarrow \infty} & \mathbb{P}\left\{  \sup_{s\in [0,t]} \left | \mathbb{E}_{\bm{X}}(\widehat{\mathrm{MCF}}^{(*)}(s;\bm{X})) - \mathbb{E}_{\bm{X}}(\mathrm{MCF}(s;\bm{X})) \right | > \epsilon \right\} \\
& = \lim_{N\rightarrow \infty} \mathbb{P}\left\{  \sup_{s\in [0,t]} \left | \frac{1}{B}\sum_{b=1}^{B}\mathbb{E}_{\bm{X}}(\widehat{\mathrm{MCF}}^{(b)}(s;\bm{X})) - \mathbb{E}_{\bm{X}}(\mathrm{MCF}(s;\bm{X})) \right | > \epsilon \right\} \\
& \leq \lim_{N\rightarrow \infty} \mathbb{P}\left\{ \frac{1}{B}\sum_{b=1}^{B}  \sup_{s\in [0,t]} \left| \mathbb{E}_{\bm{X}}(\widehat{\mathrm{MCF}}^{(b)}(s;\bm{X})) - \mathbb{E}_{\bm{X}}(\mathrm{MCF}(s;\bm{X})) \right | > \epsilon \right\}.
\end{split}
\end{equation}

Hence, it is sufficient to show that, for any bootstrap sample $b$, we have
\begin{equation} \label{eq:consistency1}
\lim_{N^{(b)}\rightarrow \infty} \mathbb{P}\left\{  \sup_{s\in [0,t]} \left | \mathbb{E}_{\bm{X}}(\widehat{\mathrm{MCF}}^{(b)}(s;\bm{X})) - \mathbb{E}_{\bm{X}}(\mathrm{MCF}(s;\bm{X})) \right | > \epsilon \right\} = 0
\end{equation}
where $N^{(b)}$ is the sample size of the bootstrap sample $b$.

Let $\eta_i$ is an indicator if sample $i$ experiences at least one failure before the censoring time $c_i$, i.e., $\eta_i=1$ if $d_i(c_i)>0$, and $\eta_i=0$ otherwise. Then, the law of large numbers lead to the following result:
\begin{equation} \label{eq:LLN}
\frac{1}{N^{(b)}}\sum_{i=1}^{N^{(b)}}I(\bm{X}_i \in  A, \eta_i =1)  \overset{\mathrm{a.s.}}{\rightarrow} \mathbb{P}\{\bm{X} \in A, \eta=1 \} = \mathbb{P}\{\bm{X} \in A\}\mathbb{P}\{\eta=1\ | \bm{X} \in A\}. 
\end{equation}
\textcolor{black}{Note that,  $\bm{X}$ and $\eta$ are not statistically independent}. However, it is reasonable to assume that $\mathbb{P}\{\eta=1\ | \bm{X} \in A\}>0$ for any $\bm{X} \in A$ (in other words, there exists no subset $A$ of $\mathbb{X}$ such that the probability of observing failures is zero), (\ref{eq:LLN}) implies
\begin{equation} \label{eq:I}
I \left( \sum_{i=1}^{N^{(b)}}I(\bm{X}_i \in  A, \eta_i =1) \geq d_0 \right) \overset{\mathrm{a.s.}}{\rightarrow} 1.
\end{equation}

Equation (\ref{eq:I}) implies that the termination rule (i.e., a terminal node must have $d_0$ systems with at least one failure) almost surely holds for any terminal node $A$. In other words, the tree almost surely has terminal nodes for all possible discretized values of $\mathbb{X}$: 
\begin{equation} \label{eq:I2}
\widehat{\mathrm{MCF}}^{(b)}(s;\bm{X}) = \sum_{x \in \mathbb{X}}I(\bm{X}=\bm{x})\widehat{\mathrm{MCF}}_{h^{(b)}}(s) + o_p(1).
\end{equation}

It follows from (\ref{eq:I2}) that
\begin{equation} \label{eq:I3}
\begin{split}
\mathrm{sup}_{s\in [0,t]} & \left | \mathbb{E}_{\bm{X}}(\widehat{\mathrm{MCF}}^{(b)}(s;\bm{X})) - \mathbb{E}_{\bm{X}}(\mathrm{MCF}(s;\bm{X})) \right | \\
& =
\mathrm{sup}_{s\in [0,t]} \left | \mathbb{E}_{\bm{X}} \left( \sum_{x \in \mathbb{X}}I(\bm{X}=\bm{x})\widehat{\mathrm{MCF}}_{h^{(b)}}(s)  \right) - \mathbb{E}_{\bm{X}} (\mathrm{MCF}(s;\bm{X})) + o_p(1)\right| \\
& = \sum_{x \in \mathbb{X}} \mathbb{P}(\bm{X}=\bm{x}) \mathrm{sup}_{s\in [0,t]}  \left | \widehat{\mathrm{MCF}}_{h^{(b)}}(s)-\mathrm{MCF}(s;\bm{x})  \right | + o_p(1).
\end{split}
\end{equation}

\color{black}
Hence, to show the uniform convergence of $\mathbb{E}_{\bm{X}} (\widehat{\mathrm{MCF}}^{(b)}(s;\bm{X}))$, it is sufficient to show the uniform convergence of $\widehat{\mathrm{MCF}}_{h^{(b)}}(s)$ to $\mathrm{MCF}(s;\bm{x})$ at each terminal node $h^{(b)}$, which requires the theorem of \cite{Anderson1993}. 
Let $\delta^{(b)}_{\cdot}(s;\bm{x})$ be the number of systems with covariate $\bm{x}$ which are still being observed at time $s$, then, the Anderson's theorem says that, if the following two conditions are satisfied: 
\begin{equation} 
\int_{0}^{t} \frac{I[\delta^{(b)}_{\cdot}(s;\bm{x})>0]}{\delta^{(b)}_{\cdot}(s;\bm{x})} v(s;\bm{x})ds\overset{\mathrm{P}}{\rightarrow}0
\label{eq:Anderson1}
\end{equation}
and
\begin{equation} 
\int_{0}^{t} \left( 1-I[\delta^{(b)}_{\cdot}(s;\bm{x})>0] \right) v(s;\bm{x})ds\overset{\mathrm{P}}{\rightarrow}0,
\label{eq:Anderson2}
\end{equation}
then, 
\begin{equation}
\mathrm{sup}_{s\in [0,t]} \left| \widehat{\mathrm{MCF}}_{h^{(b)}}(s) - \mathrm{MCF}(s;\bm{x})\right|\overset{\mathrm{P}}{\rightarrow} 0.
\label{eq:final_consistency}
\end{equation}

Hence, it is necessary to show that: 1) conditions (\ref{eq:Anderson1}) and (\ref{eq:Anderson2}) still hold for our problem, and 2) (\ref{eq:Anderson1}) and (\ref{eq:Anderson2}) still lead to the result shown in (\ref{eq:final_consistency}). 

Firstly, to show that the two conditions (\ref{eq:Anderson1}) and (\ref{eq:Anderson2}) hold, let $n^{(b)}(\bm{x})$ be the sample size on the terminal node that contains $\bm{x}$. Then, for $s\in [0,t]$ we have,
\begin{equation}
\begin{split}
\frac{1}{n^{(b)}(\bm{x})} \delta^{(b)}_{\cdot}(s;\bm{x}) & \geq  \frac{1}{n^{(b)}(\bm{x})} \sum_{i=1}^{n^{(b)}(\bm{x})} I[ C_i \geq t,\bm{X}_i   = \bm{x}] \\ &
\overset{\mathrm{a.s.}}{\rightarrow} \mathbb{P}(\bm{X} = \bm{x})\mathbb{P}(C \geq t) > 0 
\end{split}
\end{equation}
where the second line on the right hand side is based on the assumptions that $\mathbb{P}\{C>c\}>0$ for any $c\in [0,\tau)$, and $\bm{X}$ and $C$ are independent. The inequality above implies that 
\begin{equation} 
\mathrm{inf}_{s\in [0,t]}\delta^{(b)}_{\cdot}(s;\bm{x}) \overset{\mathrm{P}}{\rightarrow} \infty.  
\end{equation}
In other words, the number of systems with covariate $\bm{x}$ which are still being observed at time $s$ approaches infinity when the total sample size goes to infinity, which is intuitively true. Since $v(\cdot;\bm{x})$ is bounded over $[0,t]$, it is noted that the result above is the sufficient condition for the two conditions (\ref{eq:Anderson1}) and (\ref{eq:Anderson2}) to be satisfied.

Secondly, to show that the conditions (\ref{eq:Anderson1}) and (\ref{eq:Anderson2}) still lead to the result shown in (\ref{eq:final_consistency}), we define:
\begin{equation} 
\begin{split}
& \widehat{\mathrm{MCF}}_{h^{(b)}}(s) = \int_{0}^{s} \delta^{(b)}_{\cdot}(u;\bm{x})d\Lambda_{h^{(b)}}(u) \\
& \mathrm{MCF}_{h^{(b)}}^{(**)}(s) = \int_{0}^{s} v(u;\bm{x})I[\delta^{(b)}_{\cdot}(u;\bm{x})>0]du\\
& \mathrm{MCF}(s;\bm{x}) = \int_{0}^{s}v(u;\bm{x})du
\end{split}
\end{equation}
where $\Lambda_{h^{(b)}}(u)$ denotes the number of failures in the time interval $(0,u]$ on the tree node $h$ for the bootstrap sample $b$. 
Then, invoking the Lenglart's inequality (see (2.5.18) of \cite{Anderson1993}), for any $\eta_1, \eta_2>0$, we have
\begin{equation} \label{eq:Lenglart}
\begin{split}
\mathbb{P} & \left( \mathrm{sup}_{s\in [0,t]}   | \widehat{\mathrm{MCF}}_{h^{(b)}}(s) - \mathrm{MCF}_{h^{(b)}}^{(**)}(s) | > \eta_1 \right)  \\ 
& \leq \frac{\eta_2}{\eta_1^2} + \mathbb{P} \left( \langle \widehat{\mathrm{MCF}}_{h^{(b)}}(s) - \mathrm{MCF}_{h^{(b)}}^{(**)}(s) \rangle(t) > \eta_2 \right) \\
\end{split}
\end{equation}
where $\widehat{\mathrm{MCF}}_{h^{(b)}}(s) - \mathrm{MCF}_{h^{(b)}}^{(**)}(s)$ is a local square integrable martingale, and it follows from condition (\ref{eq:Anderson1}) and (4.1.5) of \cite{Anderson1993} that
\begin{equation} 
\langle \widehat{\mathrm{MCF}}_{h^{(b)}}(s) - \mathrm{MCF}_{h^{(b)}}^{(**)}(s) \rangle(t) = \int_{0}^{t} \frac{I[\delta^{(b)}_{\cdot}(s;\bm{x})>0]}{\delta^{(b)}_{\cdot}(s;\bm{x})} v(s;\bm{x})ds \overset{\mathrm{P}}{\rightarrow}0 .
\end{equation}

Hence, the Lenglart's inequality (\ref{eq:Lenglart}) implies that 
\begin{equation} 
\mathrm{sup}_{s\in [0,t]}   | \widehat{\mathrm{MCF}}_{h^{(b)}}(s) - \mathrm{MCF}_{h^{(b)}}^{(**)}(s) | \overset{\mathrm{P}}{\rightarrow}0.
\nonumber
\end{equation}

Since $| \mathrm{MCF}_{h^{(b)}}^{(**)}(s) - \mathrm{MCF}(s;\bm{x}) | = \int_{0}^{s} ( 1-I[\delta^{(b)}_{\cdot}(u;\bm{x})>0] ) v(u;\bm{x})du\overset{\mathrm{P}}{\rightarrow}0$ due to condition (\ref{eq:Anderson2}), we have 
\begin{equation} 
\mathrm{sup}_{s\in [0,t]} | \widehat{\mathrm{MCF}}_{h^{(b)}}(s) - \mathrm{MCF}(s;\bm{x})|\overset{\mathrm{P}}{\rightarrow} 0.	\nonumber
\end{equation}
as was to be proved. 


\color{black}
Proposition 2 gives the asymptotic variance of a single randomly drawn tree. The proof requires equation (\ref{eq:I2}) and Theorem 3.2.1 in \cite{Fleming1991}. It follows from (\ref{eq:I2})  that 
\begin{equation} \label{eq:I4}
\begin{split}
\mathrm{Var}( \sqrt{n} (\widehat{\mathrm{MCF}}(t;\bm{X}) - & \widetilde{\mathrm{MCF}}(t;\bm{X})) ) \\ & =\mathrm{Var}(\sqrt{n} \sum_{\bm{x}\in \mathbb{X}} I(\bm{X}=x)( \widehat{\mathrm{MCF}}_{h(\bm{x})}(t)  -\widetilde{\mathrm{MCF}}(t;\bm{x})  ) + o_p(1) ) \\
& =  \sum_{\bm{x}\in \mathbb{X}} \mathbb{E}( \sqrt{n}I(\bm{X}=x)(\widehat{\mathrm{MCF}}(\bm{x})(t)  -\widetilde{\mathrm{MCF}}(t;\bm{x}) ))^2  \\
& \quad - n \sum_{\bm{x}\in \mathbb{X}}I(\bm{X}=x)\mathbb{E}^2(\widehat{\mathrm{MCF}}(\bm{x})(t)  -\widetilde{\mathrm{MCF}}(t;\bm{x})) +o_p(1)
\end{split}
\end{equation} 
where $h(\bm{x})$ represents a terminal node that contains $\bm{x}$. 

Since $\widehat{\mathrm{MCF}}_{h(\bm{x})}(t)$ is asymptotically unbiased (on that particular node), the second and third term on the right-hand-side of (\ref{eq:I4}) vanishes and we have
\begin{equation}
\mathrm{Var}( \sqrt{n} (\widehat{\mathrm{MCF}}(t;\bm{X}) -  \widetilde{\mathrm{MCF}}(t;\bm{X})) ) = \sum_{\bm{x}\in \mathbb{X}} \mathbb{E}( \sqrt{\bar{n}_{\bm{x}}}(\widehat{\mathrm{MCF}}_{h(\bm{x})}(t)  -\widetilde{\mathrm{MCF}}(t;\bm{x}) ))^2 
\end{equation}  
where $\bar{n}_{\bm{x}}$ is the expected number of systems with covariates $\bm{x}$. Then, it follows from Theorem 3.2.1 of \cite{Fleming1991} that
\begin{equation}
\begin{split}
\mathrm{Var}\left( \sqrt{n} (\widehat{\mathrm{MCF}}(t;\bm{X}) - \widetilde{\mathrm{MCF}}(t;\bm{X})) \right) & = \int_{0}^{t}\pi^{-1}(s)(1-\Delta \mathrm{MCF}(s;\bm{x}))d\mathrm{MCF}(s;\bm{x})) \\
& = \sum_{x} \phi(\bm{x},t)
\end{split}
\end{equation}  
as was to be proved. 

\end{document}